\pdfoutput=1
\documentclass[sigconf]{acmart}

\usepackage{subfigure}
\usepackage{multirow}
\usepackage{multicol}
\usepackage[linesnumbered,boxed,ruled,commentsnumbered]{algorithm2e}
\usepackage{enumitem}
\setlist[itemize]{leftmargin=0.4cm}
\usepackage{pifont}
\usepackage{balance} 
\usepackage{xcolor}
\usepackage{color, colortbl}
\newcommand{\para}[1]{\vspace{2mm} \noindent \textbf{#1}}
\definecolor{crimson1}{RGB}{220, 20, 60}
\usepackage{bbm}
\definecolor{yymgray1}{HTML}{DDDFE8}
\definecolor{yymgray2}{HTML}{F2F2F2}
\definecolor{yympurple}{HTML}{AA93B4}
\definecolor{yymblue}{HTML}{E5F2FC}
\definecolor{yymblue1}{rgb}{0.6, 0.85, 0.95}
\definecolor{yymorange}{HTML}{FDECEE}
\definecolor{yymgreen}{HTML}{e6faef}
\definecolor{yymgreen1}{rgb}{0.7, 0.85, 0.85}

\setcopyright{acmlicensed}
\copyrightyear{2018}
\acmYear{2018}
\acmDOI{XXXXXXX.XXXXXXX}
\acmConference[Conference acronym 'XX]{Make sure to enter the correct
  conference title from your rights confirmation email}{June 03--05,
  2018}{Woodstock, NY}
\acmISBN{978-1-4503-XXXX-X/2018/06}

\begin{document}


\title{TopoGR: Revealing and Preserving Latent  Structure of Semantic ID in Generative Recommendation}
\author{Ziyu Zheng}
\orcid{https://orcid.org/0009-0000-3662-0832}
\affiliation{%
  \institution{Xidian University,}
    \city{Xi'an}
  \country{China}
}
\email{zhengziyu@stu.xidian.edu.cn}

\author{{Zhengshun Du}}
\affiliation{%
  \institution{Alibaba}
  \city{Beijing}
  \country{China}
}
\email{duzhengshun.dzs@alibaba-inc.com}

\author{Yaming Yang}
\orcid{https://orcid.org/0000-0002-8186-0648}
\affiliation{%
  \institution{Xidian University,}
      \city{Xi'an}
  \country{China}
}
\email{yym@xidian.edu.cn}

\author{{Bin Tong}}
 \authornote{Project leader}
\affiliation{%
  \institution{Alibaba}
  \city{Beijing}
  \country{China}
}
\email{tongbin.tb@alibaba-inc.com}

\author{Guan Wang}
\affiliation{%
  \institution{Alibaba}
  \city{Beijing}
  \country{China}
}
\email{shangfeng.wg@taobao.com}

\author{Meng Yan}
\affiliation{%
  \institution{University of Science and Technology of China}
  \city{Hefei}
  \country{China}}
\orcid{0000-0001-8478-4823}
\email{mengyan917@ustc.edu.cn}

\author{Ziyu Guan}
 \authornote{Corresponding Author}
\orcid{https://orcid.org/0000-0003-2413-4698}
\affiliation{%
  \institution{Xidian University,}
    \city{Xi'an}
  \country{China}
}
\email{zyguan@xidian.edu.cn}

\author{Wei Zhao}
\orcid{https://orcid.org/0000-0002-9767-1323}
\affiliation{%
  \institution{Xidian University,}
    \city{Xi'an}
  \country{China}
}
\email{ywzhao@mail.xidian.edu.cn}


\begin{abstract}
Semantic ID-based generative recommendation tokenizes each item into a sequence of discrete semantic IDs and predicts the next item by generating semantic IDs. However, existing methods typically regard SIDs as independent discrete symbols, while often overlooking the topology of the learned semantic ID space. We identify a structural mismatch between tokenization and generation: the tokenizer learns a structured code space with semantic neighborhood relations, whereas the generator consumes semantic ID tokens as independent categorical symbols. Consequently, item relatedness is reduced to exact semantic ID overlap, making it difficult to identify semantically similar items whose semantic IDs do not overlap. To address this issue, we propose TopoGR, a topology-preserving generative recommendation framework based on Bit-decomposable Semantic ID(Binary SID). Each Binary SID is learned in a bit-decomposable form and can be deterministically converted to a standard integer SID, while exposing an explicit Hamming geometry. TopoGR exploits this topology at three stages: binary SID features preserve Hamming proximity at the input layer; Hamming soft targets inject topology-aware supervision; and Hamming-consistent reranking aligns candidate items with the predicted binary prototype during inference. We further verify that the Hamming topology can capture item relatedness beyond exact SID matching. Experiments on four benchmark datasets show that TopoGR consistently outperforms existing state-of-the-art baselines in recommendation performance.

\end{abstract}



\maketitle

\section{Introduction}

\begin{figure}[t]
\centering
\includegraphics[width=1.0
\linewidth]{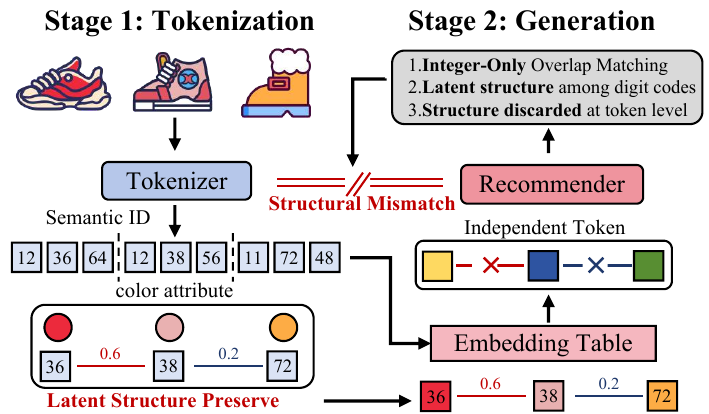}
\caption{Structural mismatch between SID tokenization and generation.
While the tokenizer preserves latent proximity among SID codes (e.g., 38 is closer to 36 than 72), integer SID lookup maps non-identical codes to independent categorical embeddings, treating both 38 and 72 as mismatches to 36 and discarding their relative proximity.}
\label{motivation_toy}
\end{figure}

Generative recommendation (GR) has recently emerged as a promising paradigm for next-item prediction~\cite{genrec_review,genrec_survey}. Rather than ranking items in the full ID space~\cite{seq_rec_survey}, GR represents each item with a sequence of discrete semantic IDs (SIDs) and casts recommendation as SID sequence generation. As illustrated in Figure~\ref{motivation_toy}, the GR pipeline is typically composed of two stages. The first stage is tokenization, which converts pretrained item semantic representations into discrete codeword sequences, usually via vector quantization~\cite{rqvae,opd,pd,onerec}. Each item is therefore assigned a SID sequence as its generative identifier. The second stage is generation, where a sequential recommender~\cite{transformers,vaswani2017attention} takes the user's interaction history as input and autoregressively predicts the SID sequence of the next item. By operating over a compact semantic token space, GR improves modeling flexibility and provides a discrete interface for incorporating item semantics into recommendation.

\begin{figure}[h]
\centering
\subfigure[Tokenization]{\includegraphics[width=0.48\columnwidth]{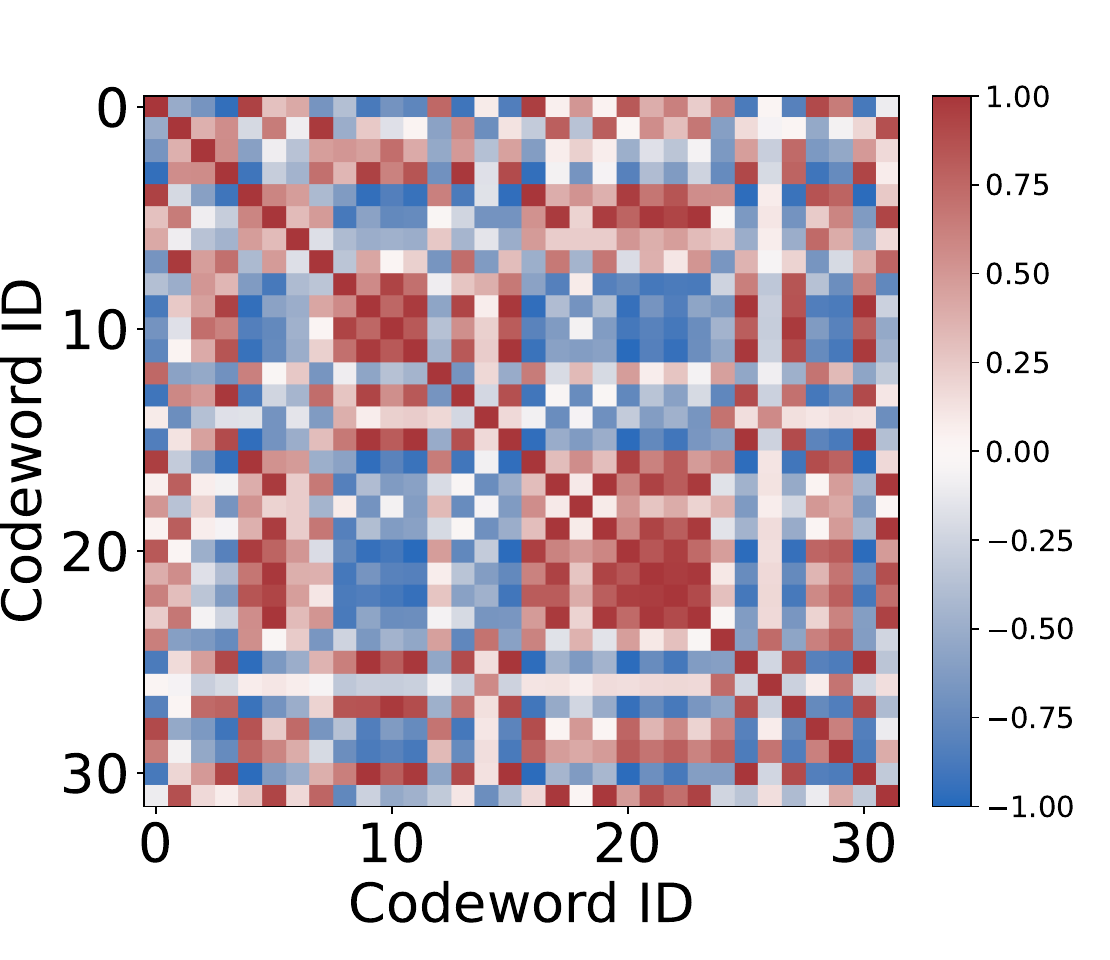}\label{fig:tok}}
\subfigure[Generation]{\includegraphics[width=0.48\columnwidth]{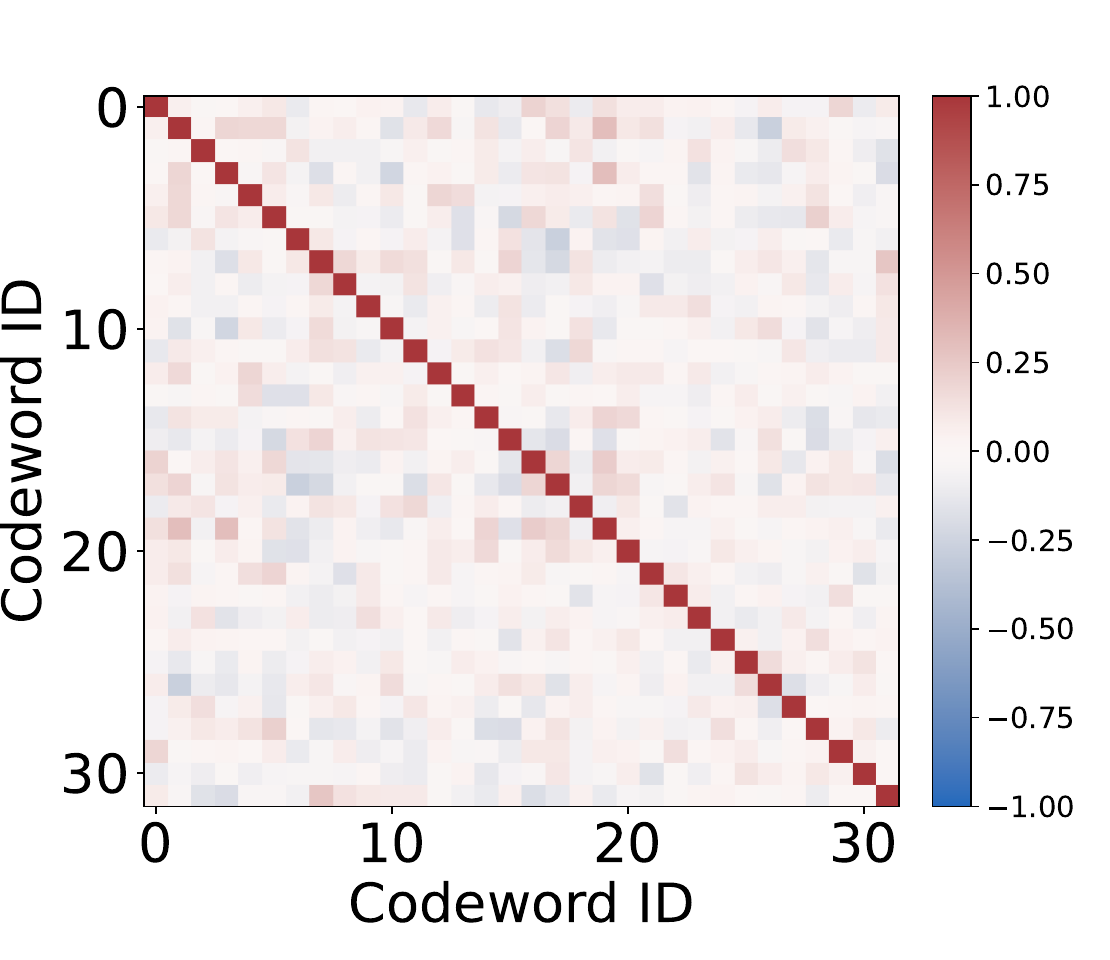}\label{fig:rec}}
\caption{Codeword similarity heatmaps for the tokenization and generation stages by OPQ tokenizer. Additional analyses for RQ-KMeans and RQ-VAE are provided in the Appendix.}
\label{fig:heatmap}
\end{figure}

Although SIDs serve as the interface between item semantics and generative models, existing GR methods~\cite{tiger, rpg, mhl} typically regard them as discrete sequences and overlook the structure within the SID space. Our analysis shows that the learned codebooks are not merely collections of independent categorical symbols, but preserve  structures inherited from the original semantic space. Figure~\ref{fig:heatmap} compares the similarity structure before and after SIDs enter the generation side, with optimized
product quantization(OPQ)~\cite{opd} used as a representative tokenizer. On the tokenization side, codeword similarities exhibit clear off-diagonal correlations, indicating latent semantic proximity among different SIDs. In contrast, on the generation side, integer SIDs are mapped to independent trainable embeddings as item inputs, whose similarity pattern is weakly structured and not explicitly aligned with the tokenizer-induced topology. This contrast indicates that the structural information encoded by the tokenizer is not explicitly exposed during generation.

This structural loss is especially problematic for item-level matching. Exact SID overlap captures only a coarse form of semantic relatedness: two items sharing SID tokens are likely to be related, but the converse does not necessarily hold. Semantically close items may fall into neighboring yet distinct codeword regions and therefore receive non-overlapping SIDs. Thus, item relatedness is determined not only by shared SID tokens, but also by the latent proximity among different codewords. As illustrated in Figure~\ref{motivation_toy}, in the tokenizer-induced latent space, two non-identical SID codes may still be close to each other, indicating that they encode similar semantic factors. However, under exact integer-token matching, all non-identical codes are simply treated as mismatches. Once these codes are mapped to independent lookup embeddings in the generator, their relative proximity is no longer explicitly preserved. Consequently, a code that is close to the target in the tokenizer's latent space and a code that is far away become indistinguishable from the perspective of exact SID matching. We refer to this problem as the \emph{structural mismatch} between tokenization and generation, which hinders GR models from exploiting the latent topology of the SID space and from generalizing to semantically similar items with non-overlapping SIDs.


The above observations suggest that the key challenge in GR is not only how to assign discrete SIDs to items, but also how to preserve the structural information learned during tokenization when these SIDs are consumed by the generator. Ideally, SID tokens should retain the efficiency and compatibility of discrete indices while exposing their latent structural relations to the recommender. Such a representation would allow the model to go beyond exact SID overlap and capture semantic relatedness between items whose SIDs are different but structurally close.

To address this structural mismatch, we propose TopoGR, a topology-preserving generative recommendation framework built upon Bit-decomposable Semantic IDs (Binary SIDs). Instead of treating each SID token as an opaque categorical label, Binary SID decomposes each token into a deterministic binary code. This representation remains fully compatible with the conventional integer SID format used for target generation, while inducing an explicit hamming geometry over the resulting binary codes. As a result, non-identical SID tokens can still be partially related through shared binary factors, providing a finer-grained notion of similarity beyond exact token overlap.

Built on Binary SIDs, TopoGR makes the SID topology available throughout the generation process. At the input stage, historical items are represented by binary SID features, allowing the model to directly access structural relations among item SIDs. During training, we introduce Hamming-aware supervision, which assigns softer penalties to predictions closer to the target code in the binary space instead of treating all incorrect codes equally. During inference, we refine the original SID generation scores by measuring the consistency between candidate Binary SIDs and the predicted binary prototype. These components follow a unified principle: 
\emph{the structure learned during tokenization should remain available to guide generation, rather than being discarded after discretization}.

To summarize, our main contributions are as follows:
\begin{itemize}
    \item We reveal a structural mismatch in current GR methods: the tokenizer learns structured SID codebooks, whereas the generator consumes SID tokens as independent categorical symbols and thus fails to exploit their latent topology.

    \item We propose TopoGR, a topology-preserving GR framework based on Bit-decomposable Semantic IDs, which retain the integer SID format while making Hamming geometry explicit.

    \item We incorporate SID topology into generation through structure aware input, Hamming aware supervision, and Hamming consistent rerank, enabling the model to capture relations between items with non-overlapping but structurally close SIDs.

    \item Extensive experiments on four benchmark datasets show that TopoGR achieves consistent improvements over competitive item ID-based and semantic ID-based recommendation methods.
\end{itemize}





\section{Related Work}
\label{sec:related}

Discriminative recommendation methods~\cite{rec_survey} formulate item prediction as a ranking problem over a fixed corpus, where each item is assigned an atomic, randomly initialized ID embedding~\cite{sasrec,bert4rec}. This paradigm suffers from limitations: the sparsity and semantic vacuity of random IDs hinder cold-start generalization and fail to capture semantic relationships among items~\cite{genrec_survey}. Generative recommendation addresses these issues by first encoding item content, such as text, images, or multimodal attributes, into continuous semantic embeddings, and then discretizing them into semantic IDs that reflect high-level semantic features. A generative model then takes the SID sequences of historically interacted items as input and generates the SID of the target item~\cite{tiger,rpg}.

Early studies, such as TIGER~\cite{tiger}, employ RQ-VAE~\cite{rqvae} for hierarchical quantization and generate SIDs in an autoregressive manner. LETTER~\cite{letter} improves the quality of quantization; LC-Rec~\cite{lc-rec} aligns the semantic and collaborative spaces through a contrastive loss; OneRec~\cite{onerec} unifies retrieval, ranking, and generation within a single architecture; and HSTU~\cite{hstu} demonstrates the scaling laws of generative recommendation models. MHL~\cite{mhl} introduces entropy-guided masked history reconstruction and curriculum learning to improve user intent modeling. Recent advances further introduce parallel generation techniques to accelerate inference in generative recommendation: RPG~\cite{rpg} enables parallel prediction of long SIDs via graph-based decoding, while LLaDA-Rec~\cite{llada-rec} and DiffGRM~\cite{diffgrm} adopt discrete diffusion for bidirectional SID generation, thereby mitigating error accumulation in autoregressive decoding.

However, these integer-SID-based methods independently embed each ID through an embedding table during recommendation. As a result, the relationships between different items are determined only by the number of shared SIDs, while the latent structural dependencies among distinct IDs are largely overlooked.

\section{Preliminary}
\subsection{Problem Formulation}
Let $\mathcal{I}$ denote the item corpus. For a user $u$, the historical interaction sequence is denoted as $\mathcal{S}_u = [v_1, v_2, \ldots, v_T], \quad v_t \in \mathcal{I}$. The goal of next-item recommendation is to predict the next item $v_{T+1}$ conditioned on $\mathcal{S}_u$:
\begin{equation}
    \hat{v}_{T+1} = \arg\max_{v \in \mathcal{I}} P(v \mid \mathcal{S}_u).
\end{equation}
In semantic ID-based generative recommendation, each item $v$ is represented by a sequence of discrete semantic tokens:
\begin{equation}
    \mathrm{SID}(v) = \mathbf{s}_v = (s_v^1, s_v^2, \ldots, s_v^M),
\quad s_v^m \in \{0,1,\ldots,K-1\},
\end{equation}
where $M$ is the number of SID positions and $K$ is the vocabulary size of each position. 



\subsection{Lookup-Free-Quantization}
Lookup-Free Quantization (LFQ)~\cite{lfq,fsq} refers to a family of discrete quantization methods that do not rely on explicit codebook lookup, which replaces explicit codebook lookup with element-wise binary quantization. In conventional vector quantization, a latent vector is assigned to one of $K$ learnable codewords in a codebook $\mathcal{C} \in \mathbb{R}^{K \times d}$. LFQ removes this lookup operation by representing each codeword as a binary vector.

Let $K = 2^r$, where $r$ is the number of bits used to represent one SID token. Given a latent vector $\mathbf{z} \in \mathbb{R}^r$, LFQ quantizes each dimension independently:
\begin{equation}
q(\mathbf{z_l}) = \operatorname{sign}(\mathbf{z}_l) =
\begin{cases}
+1, & \mathbf{z}_l > 0, \\
-1, & \mathbf{z}_l \leq 0,
\end{cases}
\quad l=1,2,\ldots,r.
\end{equation}
The resulting binary code is $\mathbf{b} = (b^1,b^2,\ldots,b^r) \in \{-1,+1\}^r$. Since there are $2^r$ possible binary patterns, the binary code $\mathbf{b}$ is equivalent to a categorical code from a vocabulary of size $K$, but without storing an explicit codebook.

\section{Methodology}
\label{sec:method}
\begin{figure*}[t]
\centering
\includegraphics[width=1.0
\linewidth]{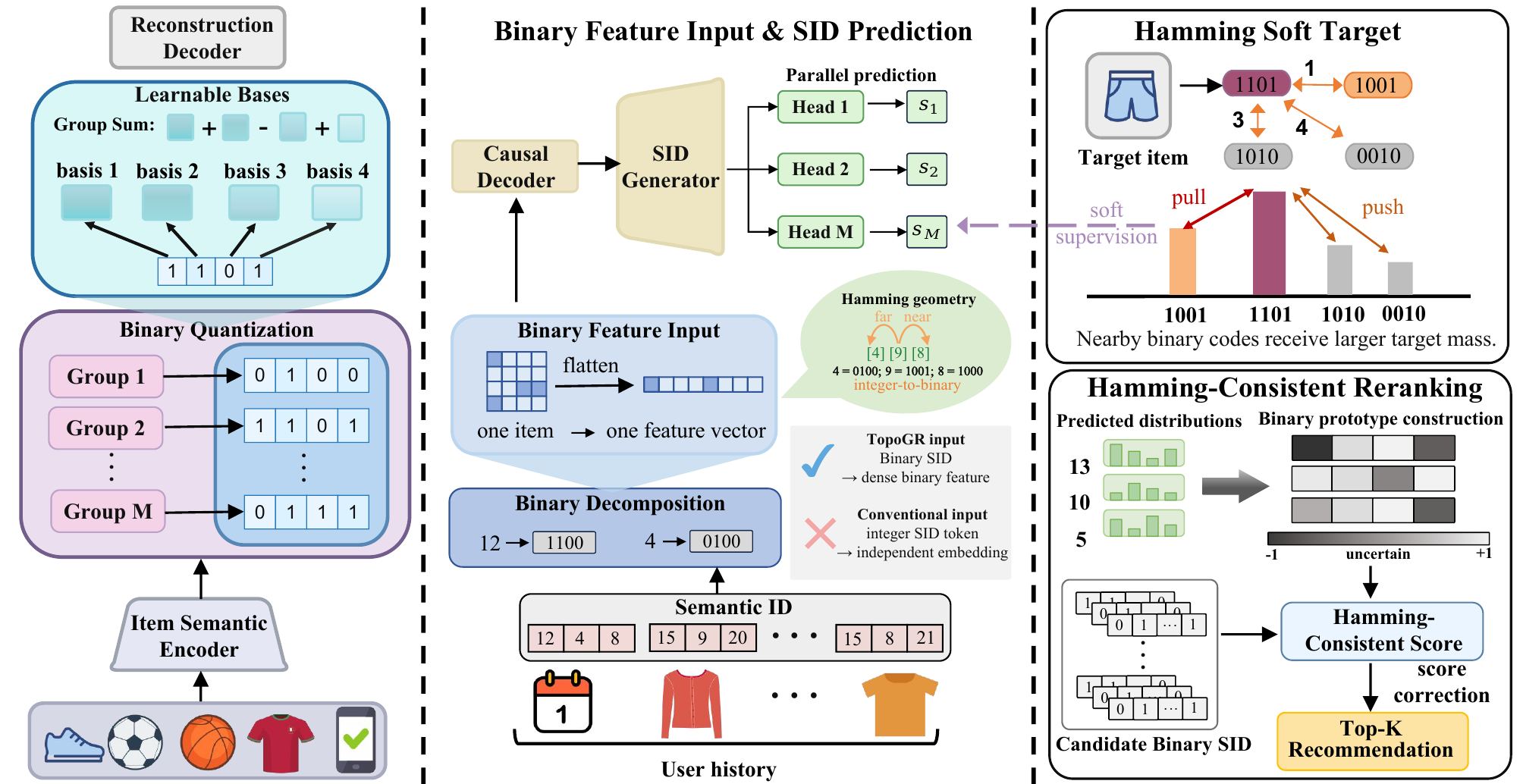}
\caption{Overview of TopoGR. TopoGR first constructs Bit-decomposable Semantic IDs through binary-structured tokenization, then uses Binary SID features as item-level inputs for generative recommendation. During training, Hamming soft targets provide topology-aware supervision, and during inference, Hamming-consistent reranking refines candidate items according to the predicted binary prototype. For illustration, binary codes are shown in $\{0,1\}$, while the model uses sign codes in $\{-1,+1\}$.}
\label{model}
\end{figure*}

To bridge SID tokenization and generation, we propose TopoGR, a topology-preserving framework for generative recommendation, consisting of bit-decomposable item tokenization, binary-feature sequence modeling, and Hamming-aware prediction and reranking, as illustrated in Figure ~\ref{model}.

\subsection{Bit-Decomposable Quantizer}
\label{sec:tokenizer}

Semantic ID-based generative recommenders represent each item \(v\) by a tuple of discrete codes. Although these codes are learned from item semantic representations, existing quantization methods usually treat them as opaque categorical labels during generation.
As a result, the generator can only exploit exact token matching, i.e., whether two items share the same SID code at a position, while the potential relation between different codes is ignored.

To preserve the fine-grained structure of the semantic code space, we introduce the Bit-Decomposable Quantizer (BDQ), which constructs topology-aware Semantic IDs for generative recommendation. BDQ follows the lookup-free quantization principle and uses multiple LFQ-style codebooks to organize an item representation into $M$ binary groups. Each group corresponds to one SID position and is quantized into an $r$-bit binary code, which can be deterministically converted into an integer SID token. Different from vanilla LFQ, BDQ further introduces a \textbf{bit-compositional reconstruction} mechanism, where each SID-position representation is explicitly composed from its bit-level bases. This design ties Hamming proximity to representation similarity and makes the learned binary topology meaningful for recommendation.


For each item $v$, let $\mathbf{x}_v \in \mathbb{R}^{d_x}$ denote its continuous semantic representation. BDQ first maps $\mathbf{x}_v$ into an $M \times r$ latent representation:
\begin{equation}
   \mathbf{Z}_v = \operatorname{Encoder}(\mathbf{x}_v) \in \mathbb{R}^{M \times r},
\end{equation}
where $M$ is the number of SID positions and $r$ is the number of bits for each SID token. Following the LFQ principle introduced in Section~3.2, BDQ treats the $M$ groups of $\mathbf{Z}_v$ as $M$ LFQ-style codebooks and quantizes each group into an $r$-bit binary code:
\begin{equation}
    \mathbf{b}_v^m = \operatorname{sign}(\mathbf{z}_v^m),
    \quad
    \mathbf{b}_v^m \in \{-1,+1\}^{r},
    \quad m=1,2,\ldots,M.
\end{equation}
The resulting Binary SID is denoted as
    $\mathbf{B}_v = [\mathbf{b}_v^1; \mathbf{b}_v^2; \ldots; \mathbf{b}_v^M]
    \in \{-1,+1\}^{M \times r}.
$
Each binary group can be deterministically converted into a conventional integer SID token:
\begin{equation}
    s_v^m =
    \sum_{l=1}^{r}
    \mathbbm{1}[b_v^{m,l} > 0] \cdot 2^{r-l},
    \quad
    s_v^m \in \{0,1,\ldots,2^r-1\}.
\end{equation}
Thus, BDQ produces both the standard integer SID sequence
$\mathbf{s}_v=(s_v^1,\ldots,s_v^M)$ and its binary form $\mathbf{B}_v$.
The former keeps compatibility with existing SID-based generative recommendation, while the latter exposes the Hamming geometry used by TopoGR.

\para{Bit-compositional Reconstruction.}
Although the above quantization follows the LFQ-style binary coding process, simply obtaining binary codes does not guarantee that their Hamming distance is semantically meaningful. To explicitly connect bit-level similarity with representation similarity, BDQ introduces a bit-compositional reconstruction mechanism.

For the $m$-th SID position, we maintain learnable base embeddings
$\{\mathbf{a}_{m,l}\}_{l=1}^{r}$, where $\mathbf{a}_{m,l}\in\mathbb{R}^{d_b}$.
Given the binary code $\mathbf{b}_v^m$, the representation of this SID position is composed as:
\begin{equation}
    \mathbf{g}_v^m =
    \sum_{l=1}^{r}
    b_v^{m,l}\mathbf{a}_{m,l}.
\end{equation}
Stacking all SID-position representations gives $\mathbf{G}_v =
    [\mathbf{g}_v^1;
     \mathbf{g}_v^2;
     \ldots;
     \mathbf{g}_v^M]
$.
This additive bit-basis formulation ties Hamming proximity to representation similarity: two codes with a small Hamming distance share most bit bases in reconstruction, and are therefore encouraged to preserve similar semantic information.

\para{Tokenizer Training Objective.}
Since the sign operation is non-differentiable, we optimize BDQ with the straight-through estimator~\cite{ste}. In the forward pass, the discrete binary code $\mathbf{B}_v$ is used to compose SID-position representations through the bit bases; in the backward pass, gradients are passed through the continuous latent representation $\mathbf{Z}_v$.

Following the bit-compositional reconstruction defined above, the tokenizer reconstructs the original item representation as
\begin{equation}
    \hat{\mathbf{x}}_v =
    \operatorname{Decoder}
    \left(
    \mathbf{G}_v
    \right),
\end{equation}
The reconstruction objective is:
\begin{equation}
\mathcal{L}_{\mathrm{recon}}
=
\frac{1}{|\mathcal{B}|}
\sum_{v \in \mathcal{B}}
\ell_{\mathrm{recon}}(\hat{\mathbf{x}}_v,\mathbf{x}_v),
\end{equation}
where $\mathcal{B}$ is a mini-batch and $\ell_{\mathrm{recon}}(\cdot)$ can be instantiated as cosine distance or mean squared error.

To avoid code collapse and encourage balanced code utilization, we additionally adopt the standard entropy auxiliary loss from LFQ:
\begin{equation}
    \mathcal{L}_{\mathrm{ent}}
    =
    \frac{1}{M}
    \sum_{m=1}^{M}
    \left[
    \mathbb{E}_{v\in\mathcal{B}}
    H(p_{v,m})
    -
    H
    \left(
    \mathbb{E}_{v\in\mathcal{B}} p_{v,m}
    \right)
    \right],
\end{equation}
where $p_{v,m}$ denotes the LFQ soft assignment distribution of the $m$-th binary group over the implicit binary codebook. The first term encourages confident per-item assignments, while the second term encourages diverse code usage at the batch level. Please see Appendix ~\ref{app:lfq_entropy} for more specific details. The tokenizer objective is:
\begin{equation}
\mathcal{L}_{\mathrm{tok}}
=
\mathcal{L}_{\mathrm{recon}}
+
\lambda_{\mathrm{ent}}\mathcal{L}_{\mathrm{ent}}.
\end{equation}
After training, each item is assigned both a integer SID $\mathbf{s}_v$ and its corresponding Binary SID $\mathbf{B}_v$. The integer form is used as the generation target, while the binary form is used as structured input and for Hamming-aware training and inference.


\para{Why Hamming Geometry Works.}
The learned Binary SID naturally defines a Hamming distance over items:
\begin{equation}
d_H(B_u,B_v)=
\sum_{m=1}^{M}\sum_{l=1}^{r}
\mathbf{1}[b_u^{m,l}\neq b_v^{m,l}].
\end{equation}
This distance is meaningful because Binary SID is learned through the bit-compositional reconstruction in Eq.~(7), rather than assigned as arbitrary binary strings.
For two items $u$ and $v$, identical bits contribute the same learnable basis components, while only different bits change the reconstructed representation:
\begin{equation}
    \mathbf{g}_u^m-\mathbf{g}_v^m
=
\sum_{l:b_u^{m,l}\neq b_v^{m,l}}
(b_u^{m,l}-b_v^{m,l})\mathbf{a}_{m,l}.
\end{equation}
Thus, Hamming distance counts how many learned basis components are changed between two Binary SID.
Since the tokenizer reconstructs item representations from these bit-basis compositions, the learned Hamming geometry is tied to the reconstruction structure and can capture item relatedness beyond exact SID matching.
A more detailed derivation is provided in Appendix~\ref{appendix:hamming_geometry}.

\subsection{Binary Feature Input for Generative Recommendation}
\label{sec:model}
After obtaining Bit-decomposable Semantic IDs, existing SID-based generators usually map each integer SID token to a learnable embedding and aggregate the embeddings across SID positions as the item
representation. However, this converts structured SID codes into independent categorical embeddings, making the tokenizer-induced topology implicit and leaving the generator to relearn such relations
from recommendation signals.

TopoGR instead directly feeds the binary structure exposed by BDQ into
the sequence model. For an item $v$, we concatenate the binary codes of
all SID positions and use the resulting Binary SID as its item-level
input feature. In this way, the input representation is no longer an
aggregation of independent SID token embeddings, but an explicit
binary encoding of the tokenizer-induced structure. The generator can
therefore observe which bits are shared or different between items,
making the Hamming topology directly available from the input layer.

Formally, for each item $v$, we flatten its Binary SID into a vector before feeding it into the sequence model. For padding
items, we use an all-zero vector and mask them out in attention. Given
a user behavior sequence $S_u = [v_1, v_2, \ldots, v_T]$, the input
sequence is constructed as:
\begin{equation}
    \mathbf{X}_u = [\mathbf{B}_{v_1}, \mathbf{B}_{v_2}, \ldots,
\mathbf{B}_{v_T}] \in \mathbb{R}^{T \times d},
\end{equation}
where $d = M r$. The sequence is then fed into a causal Transformer
decoder~\cite{gpt2}:
\begin{equation}
\mathbf{H}_u = \mathrm{CausalDecoder}(\mathbf{X}_u).
\end{equation}



This design keeps each historical item as one timestep in the user sequence, while replacing token-embedding aggregation with explicit binary SID features. 
As a result, TopoGR preserves the original item-level sequential modeling form and directly exposes the tokenizer-induced bit-level structure to the generator. 
For two items $u$ and $v$, their Binary SIDs satisfy
\begin{equation}
    \langle \mathbf{B}_u, \mathbf{B}_v \rangle_F
    =
    Mr - 2d_H(\mathbf{B}_u,\mathbf{B}_v),
    \label{eq:sign_inner_product_hamming}
\end{equation}
where $\langle\cdot,\cdot\rangle_F$ denotes the Frobenius inner product. 
Thus, before any learned transformation, Hamming-close items are already close in the input space, whereas this structural relation would be lost if SID tokens were first mapped to independent embedding vectors.


\subsection{Parallel SID Prediction}
Autoregressive SID generation predicts the tokens of the next item one by one, which couples the inference cost with the SID length. Following recent parallel SID generation methods, we instead formulate next-item generation as a multi-token prediction problem~\cite{mtp}: given the user sequence representation, the model predicts all SID positions of the next item simultaneously.

Given the binary feature sequence, the decoder produces contextualized hidden states \(\mathbf{H}=[\mathbf{h}_1,\ldots,\mathbf{h}_t]\).
Following parallel SID generation methods, we predict all SID positions of the next item simultaneously.
For each prediction position \(t\), the hidden state \(\mathbf{h}_t\) is used to predict the SID of the next item \(v_{t+1}\).
For each SID position $m$, an independent projection head maps the sequence representation $\mathbf{h}_t$ to a position-specific prediction state $\mathbf{o}_t^m$. The prediction probability over codes at position $m$ is computed by matching $\mathbf{o}^m_t$ with the corresponding code:
\begin{equation}
    P_m(k \mid \mathcal{S}_{u, \le t})
    =
    \frac{
    \exp\left(
    \operatorname{sim}(\mathbf{o}_t^m,\mathbf{e}_k^m)/\tau
    \right)
    }{
    \sum_{j=0}^{K-1}
    \exp\left(
    \operatorname{sim}(\mathbf{o}_t^m,\mathbf{e}_j^m)/\tau
    \right)
    },
    \label{eq:mtp_prob}
\end{equation}
where $\mathbf{e}_k^m$ is the embedding of the $k$-th code at the $m$-th SID position, and $\tau$ is the temperature. We keep a standard categorical prediction head over integer SID tokens to maintain compatibility with existing generative recommendation objectives. Under the standard conditional independence assumption over SID positions, the next-SID likelihood is factorized as:
\begin{equation}
    P(\mathbf{s}_{v_{t+1}}\mid \mathcal{S}_{u,\le t})
    =
    \prod_{m=1}^{M}
    P_m(s_{v_{t+1}}^m\mid \mathcal{S}_{u,\le t}).
\end{equation}
The multi-token prediction loss is then
\begin{equation}
    \mathcal{L}_{\mathrm{MTP}}
    =
    -
    \frac{1}{M}
    \sum_{m=1}^{M}
    \log
    P_m(s_{v_{t+1}}^m\mid \mathcal{S}_{u,\le t}),
    \label{eq:mtp_ce}
\end{equation}
where $s_{v_{t+1}}^m$ is the ground-truth codeword index at the m-th position. This prediction module keeps the standard semantic-token supervision used in generative recommendation. The difference is that these integer targets are derived from bit-decomposable Semantic IDs, whose binary structure will be further exploited by the Hamming-aware objectives and decoding strategy introduced next.

\subsection{Hamming Soft Targets}
The multi-token cross-entropy loss treats the target SID code as a one-hot label.
For a ground-truth code \(y\), all incorrect codes \(k\neq y\) are penalized equally, regardless of whether \(k\) differs from \(y\) by one bit or by many bits.
This is inconsistent with the bit-decomposable SID space, where Hamming-neighbor codes share more binary factors and should be considered closer semantic alternatives.

To inject this topology into supervision, we construct a Hamming-aware soft target distribution for each local SID code.
Since \(K=2^r\), every integer code \(k\in\{0,\ldots,K-1\}\) has a deterministic sign-bit representation $\mathbf{c}_k\in\{-1,+1\}^{r}$.
We pre-compute a code-level Hamming distance table:
\begin{equation}
    \mathbf{D}_{i,j}
    =
    d_H(c_i,c_j)
    =
    \sum_{l=1}^{r}
    \mathbbm{1}[c_i^l\neq c_j^l],
    \quad
    i,j\in\{0,\ldots,K-1\}.
    \label{eq:hamming_table}
\end{equation}
This table is shared by all SID positions. For a ground-truth code \(y=s_{v_{t+1}}^m\), we define the Hamming soft target over candidate codes as
\begin{equation}
    q_y(k)
    =
    \frac{
    \exp\left(-\mathbf{D}_{y,k}/\tau_H\right)
    }{
    \sum_{j=0}^{K-1}
    \exp\left(-\mathbf{D}_{y,j}/\tau_H\right)
    },
    \quad
    k=0,\ldots,K-1,
    \label{eq:hamming_soft_target}
\end{equation}
where \(\tau_H\) is the Hamming temperature.
A smaller \(\tau_H\) makes \(q_y\) close to a one-hot label, while a larger \(\tau_H\) assigns more probability mass to Hamming-neighbor codes. Given the predicted distribution \(P_m(\cdot\mid \mathcal{S}_{u,\le t})\), we minimize the KL divergence from the Hamming soft target to the model prediction:
\begin{equation}
    \mathcal{L}_{\mathrm{Ham}}
    =
    \frac{1}{M}
    \sum_{m=1}^{M}
    \operatorname{KL}
    \left(
    q_{s_{v_{t+1}}^m}(\cdot)
    \,\Vert\,
    P_m(\cdot\mid \mathcal{S}_{u,\le t})
    \right).
    \label{eq:hamming_kl}
\end{equation}
The final training objective is:
\begin{equation}
    \mathcal{L}_{\mathrm{rec}}
    =
    \mathcal{L}_{\mathrm{MTP}}
    +
    \lambda_{\mathrm{Ham}}
    \mathcal{L}_{\mathrm{Ham}}.
    \label{eq:rec_loss}
\end{equation}
The MTP term preserves exact SID prediction, while the Hamming soft target provides graded supervision among non-target codes.
Therefore, predictions close to the ground-truth code in the binary space are encouraged more than far-away codes, enabling the generator to learn the local topology of the Binary SID space instead of relying only on hard token matching.

\subsection{Inference with Hamming-Consistent Reranking}

Inference first converts the parallel token predictions into item-level scores by gathering the log-probabilities of each candidate item's SID tokens. For a candidate item $v$ with SID $\mathbf{s}_v=(s_v^1,\ldots,s_v^M)$, the original score is computed by gathering the corresponding token log-probabilities and averaging them across SID positions:
\begin{equation}
    \operatorname{Score}_{\mathrm{ori}}(v\mid\mathcal{S}_u)
=
\frac{1}{M}\sum_{m=1}^{M}\log P_m(s_v^m\mid\mathcal{S}_u).
\end{equation}
This score can be efficiently computed for all items through a vectorized gather operation over the pre-computed item-SID table.
We first select the top-\(P\) items according to \(\operatorname{Score}_{\mathrm{ori}}\) as the candidate pool \(\mathcal{C}_P\).
Although \(\operatorname{Score}_{\mathrm{ori}}\) evaluates the likelihood of each integer SID code, it still relies on exact code matching.
To further exploit the binary topology, we construct a predicted full-SID prototype from the model's output distributions.
For each SID position \(m\), let \(\mathbf{c}_k\in\{-1,+1\}^{r}\) denote the sign-bit representation of code \(k\). The expected sign vector at position \(m\) is:
\begin{equation}
    \bar{\mathbf{b}}^m
    =
    \sum_{k=0}^{K-1}
    P_m(k|\mathcal{S}_u)\mathbf{c}_k,
\end{equation}
where $\bar{\mathbf{b}}^m\in[-1,1]^r$ is a soft binary prototype. Concatenating all positions yields the predicted full-SID prototype $\bar{\mathbf{B}}
=
[\bar{\mathbf{b}}^1;\bar{\mathbf{b}}^2;\ldots;\bar{\mathbf{b}}^M]
\in\mathbb{R}^{M\times r}$. For each candidate $v\in\mathcal{C}_P$, let $\mathbf{B}_v\in\{-1,+1\}^{M\times r}$ be its Binary SID. We measure its consistency with the predicted prototype by normalized sign similarity:
\begin{equation}
    \operatorname{Sim}_{\mathrm{pred}}(v)
    =
    \frac{1}{Mr}
    \sum_{m=1}^{M}
    \sum_{l=1}^{r}
    \mathbf{\bar{b}}^{m,l} \mathbf{b}_v^{m,l}.
    \label{eq:pred_sid_sim}
\end{equation}
This similarity favors candidates whose Binary SIDs are close to the model's expected binary pattern, providing a Hamming-aware correction to the original token likelihood score. The final reranking score is computed as:
\begin{equation}
    \operatorname{Score}(v\mid\mathcal{S}_u)
=
\operatorname{Score}_{\mathrm{ori}}(v\mid\mathcal{S}_u)
+
\alpha\operatorname{Sim}_{\mathrm{pred}}(v),
\end{equation}
where $\alpha$ controls the strength of the Hamming-consistent correction. Finally, we return the top-k items according to \(\operatorname{Score}(v\mid \mathcal{S}_u)\).

This decoding procedure keeps the efficient gather-and-rank paradigm of parallel SID generation. The additional reranking only operates on a small candidate pool and does not require autoregressive beam search. By using the predicted Binary SID prototype, the decoder exploits fine-grained Hamming structure among SIDs instead of relying solely on exact integer-code likelihood.





\section{Experiment}
\label{sec:experiment}
\newcommand{\ours}[1]{\textcolor{red}{\textbf{#1}}}
\newcommand{\second}[1]{\underline{#1}}

\begin{table*}[t]
\centering
\caption{Top-K recommendation performance on four datasets. The best and second-best results are highlighted in bold and underlined, respectively.}
\label{tab:main_results}
\small

\setlength{\tabcolsep}{3.2pt}
\renewcommand{\arraystretch}{1.12}
\begin{tabular}{l|cccc|cccc|cccc|cccc}
\toprule
\multirow{2}{*}{\textbf{Model}}
& \multicolumn{4}{c}{\textbf{Beauty}}
& \multicolumn{4}{c}{\textbf{Sports and Outdoors}}
& \multicolumn{4}{c}{\textbf{Toys and Games}}
& \multicolumn{4}{c}{\textbf{CDs and Vinyl}} \\
\cmidrule(lr){2-5}
\cmidrule(lr){6-9}
\cmidrule(lr){10-13}
\cmidrule(lr){14-17}
& R@5 & N@5 & R@10 & N@10
& R@5 & N@5 & R@10 & N@10
& R@5 & N@5 & R@10 & N@10
& R@5 & N@5 & R@10 & N@10 \\
\midrule
\multicolumn{17}{c}{\textit{Item ID-based}} \\
\midrule
Caser
& 0.0205 & 0.0131 & 0.0347 & 0.0176
& 0.0116 & 0.0072 & 0.0194 & 0.0097
& 0.0166 & 0.0107 & 0.0270 & 0.0141
& 0.0116 & 0.0073 & 0.0205 & 0.0101 \\

GRU4Rec
& 0.0164 & 0.0099 & 0.0283 & 0.0137
& 0.0129 & 0.0086 & 0.0204 & 0.0110
& 0.0097 & 0.0059 & 0.0176 & 0.0084
& 0.0195 & 0.0120 & 0.0353 & 0.0171 \\

HGN
& 0.0325 & 0.0206 & 0.0512 & 0.0266
& 0.0189 & 0.0120 & 0.0313 & 0.0159
& 0.0321 & 0.0221 & 0.0497 & 0.0277
& 0.0259 & 0.0153 & 0.0467 & 0.0220 \\

BERT4Rec
& 0.0203 & 0.0124 & 0.0347 & 0.0170
& 0.0115 & 0.0075 & 0.0191 & 0.0099
& 0.0116 & 0.0071 & 0.0203 & 0.0099
& 0.0326 & 0.0201 & 0.0547 & 0.0271 \\

SASRec
& 0.0387 & 0.0249 & 0.0605 & 0.0318
& 0.0233 & 0.0154 & 0.0350 & 0.0192
& 0.0463 & 0.0306 & 0.0675 & 0.0374
& 0.0351 & 0.0177 & 0.0619 & 0.0263 \\

FDSA
& 0.0267 & 0.0163 & 0.0407 & 0.0208
& 0.0182 & 0.0122 & 0.0288 & 0.0156
& 0.0228 & 0.0140 & 0.0381 & 0.0189
& 0.0226 & 0.0137 & 0.0378 & 0.0186 \\

S$^3$-Rec
& 0.0387 & 0.0244 & 0.0647 & 0.0327
& 0.0251 & 0.0161 & 0.0385 & 0.0204
& 0.0443 & 0.0294 & 0.0700 & 0.0376
& 0.0213 & 0.0130 & 0.0375 & 0.0182 \\

\midrule
\multicolumn{17}{c}{\textit{Semantic ID-based}} \\
\midrule
RecJPQ
& 0.0311 & 0.0167 & 0.0482 & 0.0222
& 0.0141 & 0.0076 & 0.0220 & 0.0102
& 0.0331 & 0.0182 & 0.0484 & 0.0231
& 0.0075 & 0.0046 & 0.0138 & 0.0066 \\

VQ-Rec
& 0.0457 & 0.0317 & 0.0664 & 0.0383
& 0.0208 & 0.0144 & 0.0300 & 0.0173
& 0.0497 & 0.0346 & 0.0737 & 0.0423
& 0.0352 & 0.0238 & 0.0520 & 0.0292 \\

HSTU
& 0.0469 & 0.0314 & 0.0704 & 0.0389
& 0.0258 & 0.0165 & 0.0414 & 0.0215
& 0.0433 & 0.0281 & 0.0669 & 0.0357
& 0.0417 & 0.0275 & 0.0638 & 0.0346 \\

TIGER
& 0.0454 & 0.0321 & 0.0648 & 0.0384
& 0.0264 & 0.0181 & 0.0400 & 0.0225
& 0.0521 & 0.0371 & 0.0712 & 0.0432
& 0.0492 & 0.0329 & \underline{0.0748} & 0.0411 \\

RPG
& 0.0550 & 0.0381 & 0.0809 & 0.0464
& 0.0314 & 0.0216 & 0.0463 & 0.0263
& 0.0592 & 0.0401 & 0.0869 & 0.0490
& \underline{0.0498} & \underline{0.0338} & 0.0735 & \underline{0.0415} \\

MHL
& 0.0574 & \underline{0.0424} & 0.0795 & 0.0495
& 0.0359 & \underline{0.0249} & 0.0511 & 0.0298
& \underline{0.0672} & \underline{0.0489} & \underline{0.0903} & \underline{0.0564}
& 0.0488 & 0.0337 & 0.0701 & 0.0405 \\

DiffGRM
& \underline{0.0603} & 0.0414 & \underline{0.0876} & \underline{0.0502}
& \underline{0.0363} & 0.0245 & \underline{0.0550} & \underline{0.0305}
& 0.0618 & 0.0455 & 0.0834 & 0.0524
& 0.0348 & 0.0228 & 0.0550 & 0.0293 \\

\midrule
\textbf{TopoGR}
& \textbf{0.0620} & \textbf{0.0439} & \textbf{0.0883} & \textbf{0.0522}
& \textbf{0.0374} & \textbf{0.0258} & \textbf{0.0554} & \textbf{0.0314}
& \textbf{0.0712} & \textbf{0.0497} & \textbf{0.0995} & \textbf{0.0588}
& \textbf{0.0531} & \textbf{0.0364} & \textbf{0.0783} & \textbf{0.0445} \\

\bottomrule
\end{tabular}
\end{table*}
To evaluate the effectiveness of TopoGR and validate our hypothesis that preserving SID topology benefits generative recommendation, we aim to answer the following research questions:
\begin{itemize}
    \item \textbf{RQ1(Performance):}
    How does TopoGR perform compared with state-of-the-art recommendation baselines?

    \item \textbf{RQ2(Effectiveness of Binary SID):}
    Does BDQ provide a more useful binary SID structure than conventional quantization or random binary assignment?

    \item \textbf{RQ3(Impact of Topology-aware Modeling):}
    Do binary feature input, Hamming soft targets, and Hamming-consistent reranking effectively improve generative recommendation?

    \item \textbf{RQ4(Role of Hamming Geometry):}
    Does Hamming proximity capture item relatedness beyond exact SID overlap, and does this topology help generalize to sparse or cold-start items?
\end{itemize}

\subsection{Experimental Settings}
\para{Datasets.}
We evaluate our model on four categories from the Amazon Review dataset~\cite{amzonreview2014}: Sports \& Outdoors (Sports), Beauty, Toys \& Games (Toys), and CDs \& Vinyl (CDs). Following prior work~\cite{rpg,lc-rec,tiger}, user reviews are treated as interactions and chronologically ordered to form interaction sequences. For item metadata, we concatenate title, brand, category, and description into natural language sentences to enable semantic representation learning. Appendix~\ref{sec:dataset_statistics} summarizes the statistics of all four datasets.

\para{Baselines.}
We evaluate TopoGR against both item ID-based and semantic ID-based recommendation baselines. Specifically, the item ID-based methods include Caser~\cite{caser}, GRU4Rec~\cite{gru4rec}, HGN~\cite{hgn}, BERT4Rec~\cite{bert4rec}, SASRec~\cite{sasrec}, FDSA~\cite{fsda}, and S³-Rec~\cite{s3-rec}, while the semantic ID-based methods include RecJPQ~\cite{petrov2024recjpq}, VQ-Rec~\cite{vq-rec}, HSTU~\cite{hstu}, TIGER~\cite{tiger}, RPG~\cite{rpg}, MHL~\cite{mhl}, and DiffGRM~\cite{diffgrm}. 


\para{Evaluation Protocol.}
We adopt the standard leave-one-out evaluation scheme~\cite{tiger,rpg}. Recommendation performance is measured by two widely used ranking metrics: Recall@$K$ and Normalized Discounted Cumulative Gain (NDCG@$K$), with $K \in \{5, 10\}$. All results are reported based on the best validation performance.

\para{Implementation Details.}
We use Sentence-T5-base~\cite{sentence-t5} to encode item metadata and obtain 768-dimensional item embeddings, which are then reduced to 256 dimensions via PCA. The decoder adopts the same architecture as ~\cite{rpg}. Specifically, the hidden dimension is set to the product of the bit number and the SID length. We
use a 2-layer Transformer decoder. The feed-forward dimension is set to 1024, and the number of attention heads is set to 4. The size of the candidate pool is set to 1000. We set the number of bits to 8, yielding an integer ID range from 0 to 255 for each SID position. For hyperparameter tuning, we search the learning rate in $\{0.01, 0.003, 0.001, 0.0005\}$ and the SID length in $\{8,16,32,64,128\}$. The weights of both the Hamming Soft Targets loss and the Hamming-aware correction score during inference are tuned from 0 to 1.0 with an interval of 0.1. For detailed hyperparameter experiments, please refer to Appendix~\ref{app:hyperparam}. For baseline results, we adopt the reported numbers from ~\cite{tiger}. For all other datasets and baseline models, we reproduce the results using the official implementations or the RecBole~\cite{recbole} to ensure fair comparison. All experiments were conducted on a single NVIDIA H20 GPU with 96GB memory.



\subsection{Performance Comparison(RQ1)}
Table~\ref{tab:main_results} presents the overall performance on four datasets. TopoGR achieves the best results across all datasets and metrics. Compared with RPG, which also adopts a parallel SID generation framework, TopoGR consistently improves performance by replacing the conventional SID-token embedding table with Binary SID features.
The gains are especially clear on Toys and Sports: on Toys, our method improves N@5 by about \(24\%\) over RPG; on Sports, it improves N@5 by around \(19\%\). These results indicate that the performance gain mainly comes from making the bit-level structure of SIDs visible to the generator, rather than from changes in the decoding framework.

Moreover, TopoGR outperforms recent state-of-the-art generative recommendation models such as MHL and DiffGRM without requiring additional masked reconstruction or diffusion-style generation. This suggests that explicitly preserving the internal geometry of Semantic IDs can provide a strong and lightweight
alternative to more complex generation objectives. The consistent improvements on both small item corpora such as Beauty and Toys and the larger CDs dataset further show that the proposed topology-preserving design is robust under different item-space scales.

\begin{figure}[h]
\centering
\includegraphics[width=0.9\columnwidth]{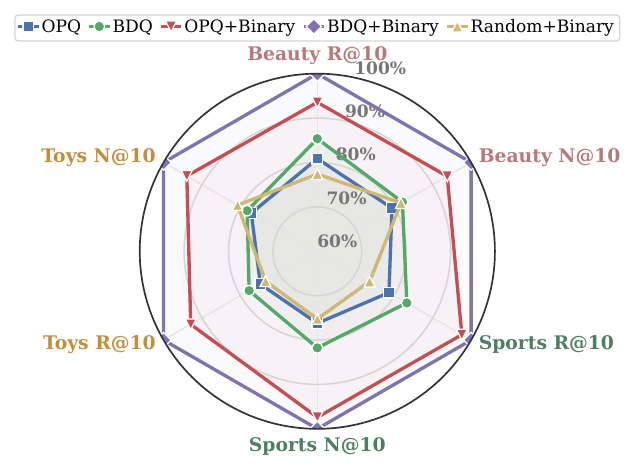}
\caption{Ablation study on Binary SID.}
\label{fig:binary_sid_ablation}
\end{figure}

\begin{figure}[h]
\centering
\includegraphics[width=1.0\columnwidth]{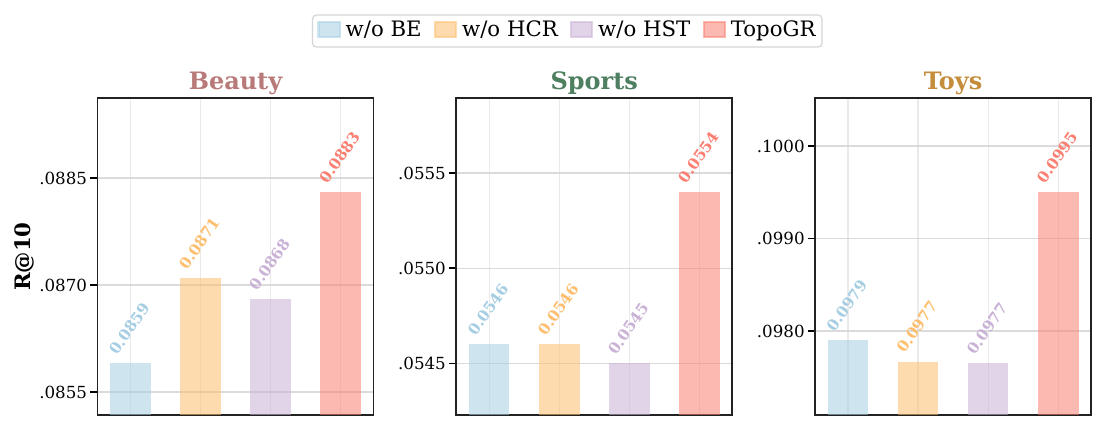}
\caption{Component ablation study.}
\label{fig:component_ablation}
\end{figure}

\subsection{Ablation Study}
We conduct ablation studies to examine whether the improvements of TopoGR come from the proposed topology-preserving design.

\para{Binary SID construction(RQ2). }
Figure~\ref{fig:binary_sid_ablation} compares different ways of introducing binary structures into semantic IDs. OPQ+Binary consistently improves over OPQ, showing that exposing a binary form of SID tokens can benefit the generator. However, its performance is still lower than BDQ+Binary, which directly learns bit-decomposable codes during tokenization. This suggests that the binary structure is more effective when it is aligned with the quantization process, rather than imposed after discrete indices have been produced. In contrast, Random+Binary is consistently inferior to topology-aware binary constructions, indicating that the improvement comes from meaningful binary topology rather than the binary representation itself.

\para{Model components(RQ3). }
Figure~\ref{fig:component_ablation} further studies the contribution of the main components in TopoGR. Removing any component leads to a consistent performance drop across datasets. The variant without BE shows that the basis embedding used in tokenization contributes to learning better binary-structured SIDs. The degradation of w/o HST indicates that treating all non-target codes equally is suboptimal; Hamming-aware supervision provides graded training signals by assigning higher tolerance to codes that are closer to the target in the binary space. The drop caused by removing HCR indicates that the predicted binary prototype provides useful information for reranking. Overall, the full model achieves the best results, demonstrating that the proposed components are complementary in preserving and exploiting SID topology.

\subsection{Further Analysis(RQ4)}
\para{Effectiveness of Hamming Proximity.}
We further investigate whether the Hamming geometry of Binary SIDs captures item relatedness beyond exact SID overlap. To isolate the effect of Hamming proximity, we control the maximum integer SID overlap between the target item and the user's historical items. For each test instance, we compute the number of shared integer SID tokens between the target item and each historical item, and group the instance by the maximum overlap count. Buckets 0--4 contain instances with exactly the corresponding maximum overlap, while the $\geq 5$ bucket collects high-overlap cases. Within each bucket, we compare instances whose closest maximum-overlap historical items are Hamming-close or Hamming-far from the target item. The detailed grouping procedure is described in Appendix~\ref{app:hamming_grouping}.

Table~\ref{tab:hamming_close_far_n10} reports the controlled comparison. Under the same integer SID overlap, Hamming-close pairs consistently obtain higher NDCG@10 than Hamming-far pairs on both Toys and Beauty. The Close/Far ratio ranges from 1.99$\times$ to 6.22$\times$. Notably, even when the integer SID overlap is zero, the Hamming-close group outperforms the Hamming-far group by 3.06$\times$ on Toys and 3.37$\times$ on Beauty. This verifies that Hamming proximity provides additional semantic signals that cannot be captured by exact SID matching alone.

\begin{figure}[h]
\centering
\subfigure[Beauty]{\includegraphics[width=0.48\columnwidth]{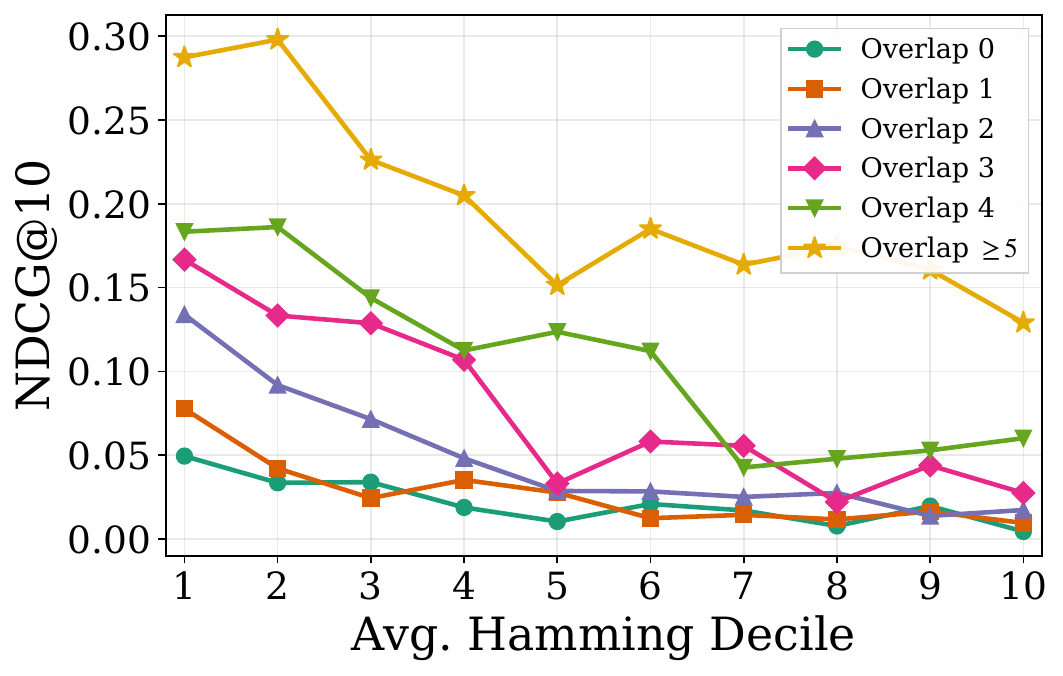}\label{fig:beauty}}
\subfigure[Toy]{\includegraphics[width=0.48\columnwidth]{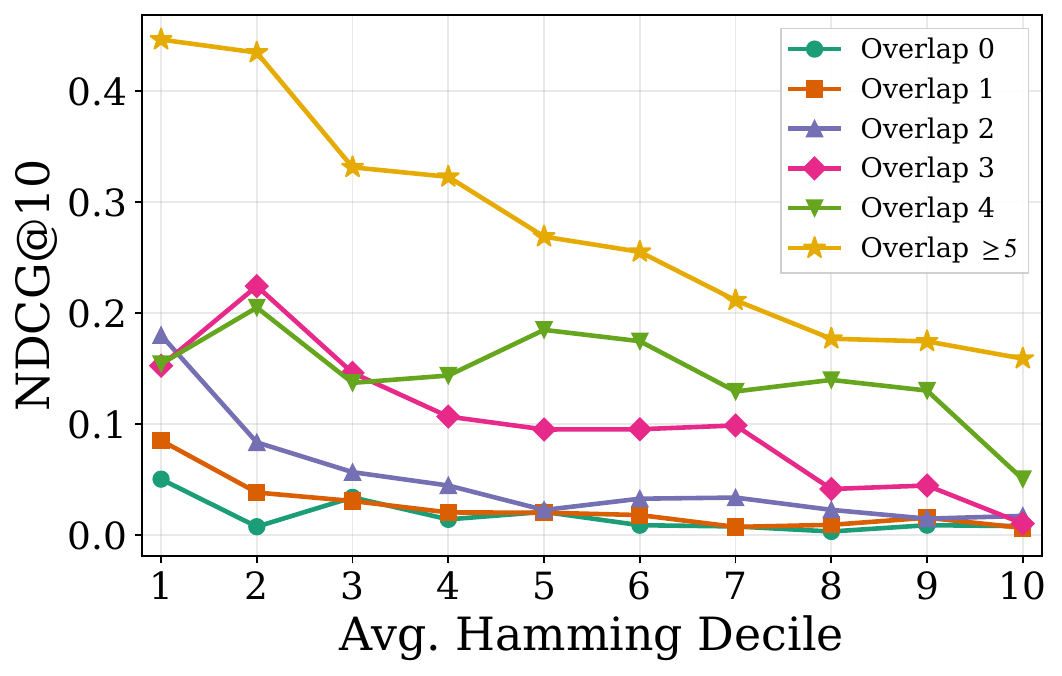}\label{fig:toy}}
\caption{Effect of Hamming distance at fixed SID-overlap.}
\label{fig:hamming_effect}
\end{figure}

Figure~\ref{fig:hamming_effect} provides a finer-grained view by sorting pairs within each overlap bucket according to their average Hamming distance. Across most overlap buckets, NDCG@10 decreases as the Hamming decile increases, indicating that items with closer binary codes are more likely to be relevant even when their integer SID overlap is fixed. These results support our central motivation: Binary SIDs expose meaningful topology among non-identical SID tokens. This topology explains why Hamming-aware supervision and Hamming-consistent decoding can improve generative recommendation beyond overlap-based SID matching. Appendix~\ref{app:bit_content} further validates this observation through content-similarity analysis under integer and Binary SID overlap.

\begin{table}[h]
\centering
\caption{
Comparison of Hamming-close and Hamming-far groups under the same SID overlap count on NDCG@10.}
\label{tab:hamming_close_far_n10}
\small
\setlength{\tabcolsep}{3pt}
\resizebox{\columnwidth}{!}{
\begin{tabular}{lccc|ccc}
\toprule
\multirow{2}{*}{\textbf{Overlap}}
&
\multicolumn{3}{c|}{\textbf{Toy}}
&
\multicolumn{3}{c}{\textbf{Beauty}}
\\
\cmidrule(lr){2-4}
\cmidrule(lr){5-7}
&
\textbf{Close}
&
\textbf{Far}
&
\textbf{Ratio}
&
\textbf{Close}
&
\textbf{Far}
&
\textbf{Ratio}
\\
\midrule
0
& 0.0338
& 0.0110
& 3.06$\times$
& 0.0305
& 0.0091
& 3.37$\times$
\\
1
& 0.0496
& 0.0122
& 4.06$\times$
& 0.0453
& 0.0152
& 2.98$\times$
\\
2
& 0.1007
& 0.0270
& 3.73$\times$
& 0.0991
& 0.0226
& 4.38$\times$
\\
3
& 0.1958
& 0.0315
& 6.22$\times$
& 0.1193
& 0.0423
& 2.82$\times$
\\
4
& 0.2085
& 0.0821
& 2.54$\times$
& 0.1457
& 0.0732
& 1.99$\times$
\\
$\geq 5$
& 0.3915
& 0.1476
& 2.65$\times$
& 0.2948
& 0.1375
& 2.14$\times$
\\
\bottomrule
\end{tabular}
}
\end{table}

\begin{figure}[h]
\centering
\includegraphics[width=1.0\columnwidth]{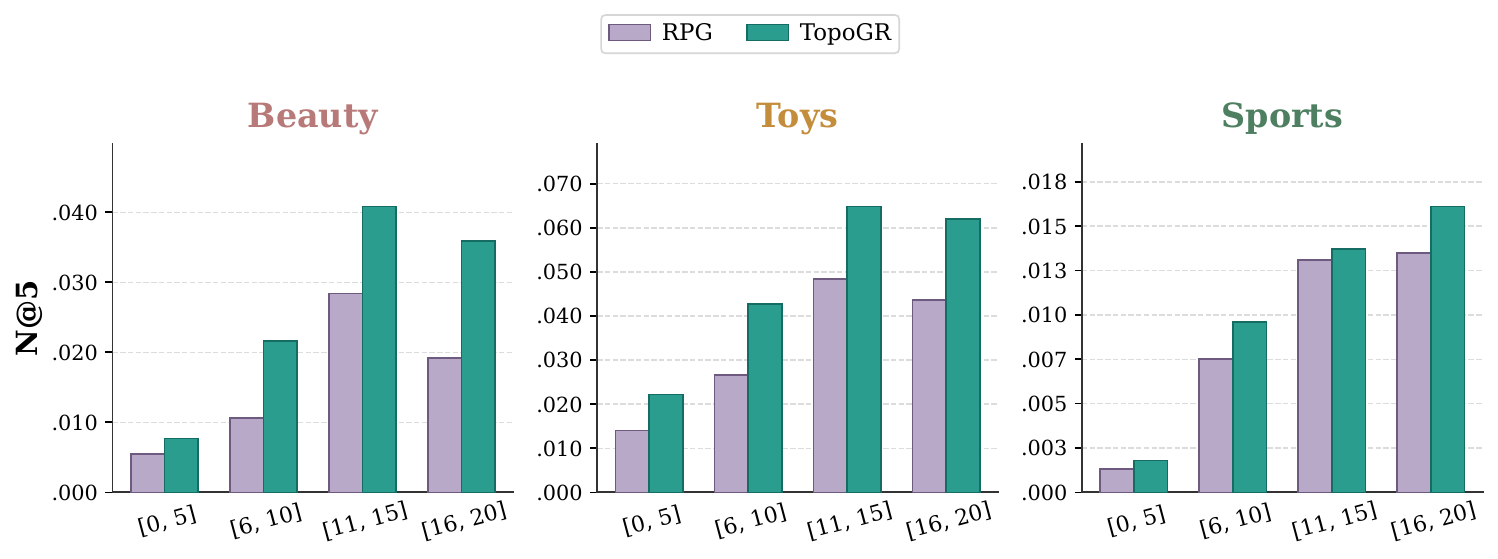}
\caption{Cold-start recommendation performance.}
\label{fig:cold_start}
\end{figure}

\para{Cold-Start Recommendation. }
We evaluate cold-start performance by grouping test cases according to the frequency of target items in the training set. As shown in Figure~\ref{fig:cold_start}, TopoGR consistently outperforms RPG across all buckets on Beauty, Toys, and Sports, with larger gains on low-frequency items where item-specific supervision is limited. These results show that preserving SID topology improves generalization to sparse items. Unlike conventional SID-based generators that mainly rely on exact token overlap, TopoGR exploits Hamming proximity between different Binary SID codes, enabling low-frequency items to benefit from semantically related neighbors in the binary SID space.

\section{Conclusion}

In this paper, we reveal a structural mismatch in semantic ID-based generative recommendation, where structured SID spaces learned by tokenizers are reduced to independent categorical tokens in generators. We propose TopoGR with Bit-decomposable Semantic IDs, which preserve the standard integer SID format while exposing Hamming geometry. TopoGR exploits this topology through binary input representation, Hamming-aware supervision, and Hamming-consistent reranking. Experiments on four datasets show consistent improvements over strong baselines. Ablation and further analyses confirm that preserving SID topology provides useful signals beyond exact SID overlap, especially for sparse and cold-start items.

\bibliographystyle{ACM-Reference-Format}
\balance
\bibliography{refs}

\appendix
\newpage
\section{More Details}
\subsection{Datasets and Hyperparameters details}
\label{sec:dataset_statistics}

\begin{table}[h]
\centering
\caption{Statistics of the processed datasets. ``Avg. $t$'' denotes the average number of interactions per input sequence.}
\label{tab:dataset_statistics}
\begin{tabular}{lrrrr}
\toprule
\textbf{Datasets} & \textbf{Users} & \textbf{Items} & \textbf{Interactions} & \textbf{Avg. $t$} \\
\midrule
\textbf{Sports} & 35,599 & 18357 & 296,337 & 8.32 \\
\textbf{Beauty} & 22,363 & 12,101 & 198,502 & 8.88 \\
\textbf{Toys}   & 19,412 & 11,924 & 167,597 & 8.63 \\
\textbf{CDs}    & 75,258 & 64,443 & 1,022,334 & 13.58 \\
\bottomrule
\end{tabular}
\end{table}

\begin{table}[h]
\centering
\caption{Detailed hyperparameter settings of TopoGR on different datasets.}
\label{tab:final_hyperparams}
\begin{tabular}{lrrrr}
\toprule
Hyperparameter & Beauty & Sports & Toys & CDs \\
\midrule
Learning Rate & 0.01 & 0.003 & 0.003 & 0.0005 \\
$n_{codebook}$ & 64 & 64 & 64 & 64 \\
Bit Number $r$ & 8 & 8 & 8 & 8 \\
Code Size $K$ & 256 & 256 & 256 & 256 \\
$\tau$ & 0.03 & 0.03 & 0.03 & 0.03 \\
$\lambda_{\mathrm{ent}}$ & 0.1 & 0.1 & 0.1 & 0.1 \\
$\lambda_{\mathrm{Ham}}$ & 0.1 & 0.1 & 0.1 & 0.1 \\
$\alpha$ & 0.3 & 0.3 & 0.3 & 0.3 \\
$\tau_H$ & 0.1 & 0.1 & 1.0 & 1.0 \\

\bottomrule
\end{tabular}
\end{table}

\begin{table}[h]
\centering
\caption{Full-SID uniqueness statistics. The collapse ratio is close to zero on all datasets, indicating that the learned full SIDs rarely collapse.}
\label{tab:sid_collapse}
\begin{tabular}{lrrr}
\toprule
\textbf{Datasets} & \textbf{Items} & \textbf{Unique SIDs} & \textbf{Collapse Ratio} \\
\midrule
\textbf{Beauty} & 12,101 & 12,099  & 0.000165 \\
\textbf{Toys}   & 11,924 & 11,915  & 0.000755 \\
\textbf{Sports} & 18,357 & 18,317  & 0.002179 \\
\textbf{CDs}    & 64,443 & 64,252  & 0.002964 \\
\bottomrule
\end{tabular}
\end{table}
\subsection{SID Collapse Analysis}
We further examine whether the learned SIDs suffer from collapse, where multiple items are assigned to the same SID. As shown in Table~\ref{tab:sid_collapse}, the learned SIDs exhibit negligible collapse across all datasets. 
The collapse ratio is consistently below 0.003.
These results indicate that the proposed bit-decomposable tokenizer can assign highly distinguishable SIDs to items and does not suffer from severe SID collapse.

\subsection{Implementation of LFQ Entropy Regularization}
\label{app:lfq_entropy}

We adopt the standard entropy regularization of Lookup-Free
Quantization (LFQ)~\cite{lfq}. This regularizer is inherited
from LFQ and is not a methodological contribution of TopoGR.

For each SID position $m$, the encoder produces a continuous
pre-quantization vector $\mathbf{z}_{v,m}\in\mathbb{R}^{r}$. LFQ
performs hard quantization independently along each dimension:
\[
\mathbf{b}_{v,m}
=
\operatorname{sign}(\mathbf{z}_{v,m})
\in\{-1,+1\}^{r}.
\]
Although LFQ does not maintain learnable codebook centroids, the
$K=2^r$ possible binary patterns can be deterministically enumerated as
\[
\mathcal{C}
=
\{\mathbf{c}_{k}\}_{k=0}^{K-1},
\qquad
\mathbf{c}_{k}\in\{-1,+1\}^{r}.
\]
These binary vertices are fixed buffers rather than learnable
parameters. They are only instantiated during training to compute the
entropy regularizer; the hard forward quantization remains an
element-wise sign operation without nearest-codeword lookup.

Given $\mathbf{z}_{v,m}$, the implementation defines a differentiable
soft assignment over the implicit binary codebook as
\[
p_{v,m}(k)
=
\frac{
\exp\left(2\beta\,\mathbf{z}_{v,m}^{\top}\mathbf{c}_{k}\right)
}{
\sum_{j=0}^{K-1}
\exp\left(2\beta\,\mathbf{z}_{v,m}^{\top}\mathbf{c}_{j}\right)
},
\qquad k=0,\ldots,K-1,
\]
where $\beta$ denotes the inverse temperature. This formulation is
equivalent, up to code-independent terms, to applying a softmax over
the negative squared Euclidean distances between
$\mathbf{z}_{v,m}$ and the fixed binary vertices. Importantly,
$p_{v,m}$ is computed from the continuous pre-quantization vector,
rather than from the discrete output of the sign operation.

For a mini-batch $\mathcal{B}$, we first compute the average assignment
distribution at SID position $m$:
\[
\bar{p}_{m}
=
\frac{1}{|\mathcal{B}|}
\sum_{v\in\mathcal{B}}p_{v,m}.
\]
The entropy regularizer is then defined as
\[
\mathcal{L}_{\mathrm{ent}}
=
\frac{1}{M}
\sum_{m=1}^{M}
\left[
\frac{1}{|\mathcal{B}|}
\sum_{v\in\mathcal{B}}
H\!\left(p_{v,m}\right)
-
\gamma H\!\left(\bar{p}_{m}\right)
\right],
\]
where
\[
H(p)=-\sum_{k=0}^{K-1}p(k)\log p(k),
\]
and $\gamma$ controls the strength of batch-level code diversity. The
first term encourages each item to make a confident assignment,
whereas the second term encourages the batch-aggregated distribution
to utilize diverse binary codes. We use $\gamma=1$ in our experiments.

The entropy path is fully differentiable with respect to
$\mathbf{z}_{v,m}$. In particular,
\[
\frac{\partial p_{v,m}(k)}
{\partial \mathbf{z}_{v,m}}
=
2\beta p_{v,m}(k)
\left(
\mathbf{c}_{k}
-
\sum_{j=0}^{K-1}
p_{v,m}(j)\mathbf{c}_{j}
\right).
\]
Thus, gradients from $\mathcal{L}_{\mathrm{ent}}$ are directly
back-propagated to the encoder through the continuous logits. The
binary vertices $\mathbf{c}_{k}$ receive no gradients because they are
fixed buffers.

Separately, the reconstruction path uses the straight-through
estimator:
\[
\widetilde{\mathbf{b}}_{v,m}
=
\mathbf{z}_{v,m}
+
\operatorname{sg}\!\left(
\mathbf{b}_{v,m}-\mathbf{z}_{v,m}
\right),
\]
where $\operatorname{sg}(\cdot)$ denotes stop-gradient. The forward
value of $\widetilde{\mathbf{b}}_{v,m}$ is the hard binary code, while
its backward derivative with respect to $\mathbf{z}_{v,m}$ is treated
as the identity. The resulting binary code is subsequently composed
with the learnable bit bases for semantic reconstruction. Therefore,
the soft assignment is used only for entropy regularization and does
not replace the hard Binary SID used by the tokenizer.



\subsection{Complexity Analysis}
Let $L$ denote the user sequence length, $M$ the number of SID positions per item, $r$ the number of bits per SID position, and $K=2^r$ the local code vocabulary size. The binary item feature dimension is $d=Mr$. Let $|\mathcal{I}|$ denote the number of items in the corpus, and let $P$ denote the size of the candidate pool used for reranking.

\para{Training.}
TIGER flattens all $M$ SID tokens of each item into the sequence. Therefore, the effective sequence length becomes $LM$, and the self-attention complexity is
$O(L^2M^2d).$ This cost grows quadratically with the SID length and becomes expensive when long SIDs are used. RPG keeps each item as one timestep and aggregates the $M$ SID tokens into a single item representation, reducing the self-attention complexity to $O(L^2d).$However, the aggregation may lose fine-grained token-level structure. TopoGR also keeps each historical item as one timestep. It represents each item as a binary feature vector of dimension
$d=Mr$, and therefore has the same self-attention complexity as RPG: $O(L^2d).$ In addition to the Transformer backbone, TopoGR uses $M$ parallel
prediction heads to compute logits over $K$ codes at each sequence position. This introduces an additional cost of $O(LMKd)$.

The Hamming soft-target loss computes a KL divergence over $K$ codes for each SID position, which costs $O(LMK).$ Therefore, the overall training complexity of TopoGR is $O(L^2d + LMKd + LMK)$.
Since the last term is dominated by the prediction-head computation, this can be simplified as
$O(L^2d + LMKd)$. The prediction-head computation is linear in the sequence length and fully parallelizable across SID positions and code indices.
Although this term can be non-negligible when $M$ and $K$ are large, it is efficiently implemented with batched matrix operations and remains practical in our experiments.

\begin{figure}[t]
\centering
\includegraphics[width=0.95\columnwidth]{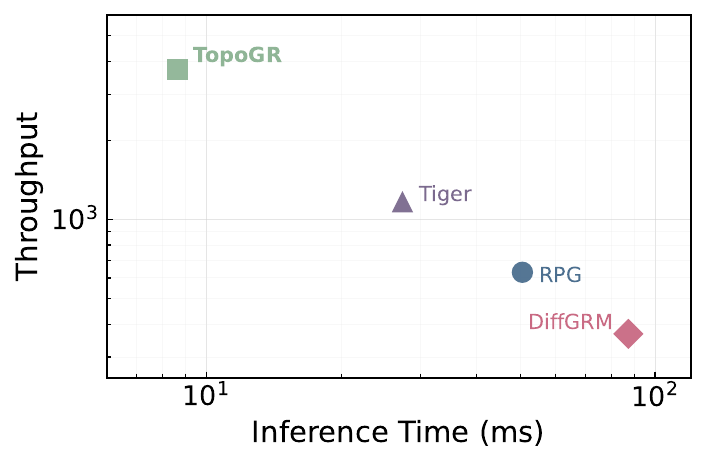}
\caption{Comparison of inference time and throughput of different baselines on the Beauty dataset.}
\label{fig:inference_efficiency}
\end{figure}

\para{Inference.}
TIGER performs autoregressive SID generation. With beam width $W$, its decoding cost scales with the SID length as $O(MWKd)$, which introduces a serial decoding bottleneck. RPG avoids serial generation by predicting SID positions in parallel, but its graph-constrained decoding requires additional candidate propagation and graph traversal, whose cost depends on the graph construction and decoding hyperparameters. TopoGR predicts all $M$ SID positions in parallel. Computing the logits of all SID positions costs $O(MKd).$
Then, for each item in the corpus, TopoGR gathers the
log-probabilities of its $M$ SID tokens and averages them to obtain the original item score. This full-corpus scoring step costs $O(|\mathcal{I}|M).$
After selecting the top-$P$ candidates according to the original score, Hamming-consistent reranking computes the normalized bit-level similarity between each candidate Binary SID and the predicted
binary prototype, which costs
$O(PMr)$. Thus, the overall inference complexity of TopoGR is $O(MKd + |\mathcal{I}|M + PMr),$ plus the cost of top-$P$ selection. With an efficient top-$P$
operator, the selection can be implemented in approximately linear time with respect to $|\mathcal{I}|$ in practice. TopoGR therefore eliminates the serial autoregressive decoding
bottleneck of TIGER and avoids the iterative graph traversal used by RPG. Its inference procedure consists of parallel SID prediction, vectorized full-corpus gather scoring, and lightweight reranking
over a small candidate pool.

Figure~\ref{fig:inference_efficiency} further compares the inference latency and throughput of different generative recommendation methods under the same experimental environment. TIGER performs autoregressive SID generation, whose sequential decoding procedure results in high latency. RPG enables parallel SID prediction but introduces
graph-constrained decoding, which incurs additional candidate propagation and graph traversal overhead. DiffGRM adopts bidirectional masked diffusion and requires multiple denoising iterations, leading to high inference latency. In contrast, TopoGR predicts all SID positions in parallel and performs full-corpus scoring and Hamming-consistent reranking with vectorized matrix operations. Although TopoGR still scores the item corpus through gather operations, this step is highly parallelizable and avoids serial decoding or iterative graph traversal. Consequently, TopoGR achieves low inference latency and high throughput.

\section{More Experiments}

\subsection{Additional Codeword Similarity Analysis}

\begin{figure}[h]
\centering
\subfigure[RQ-Kmeans]{\includegraphics[width=0.48\columnwidth]{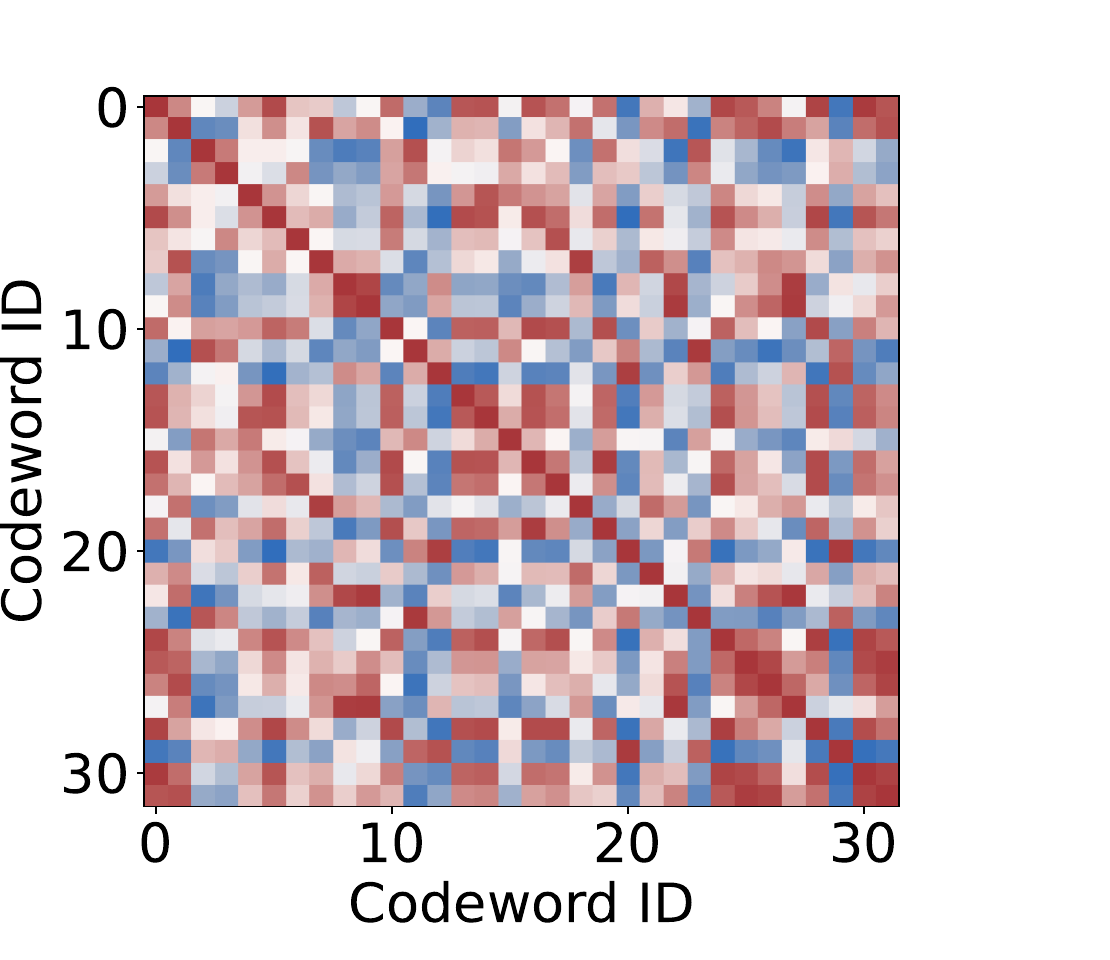}\label{fig:rqkmeans}}
\subfigure[RQ-VAE]{\includegraphics[width=0.48\columnwidth]{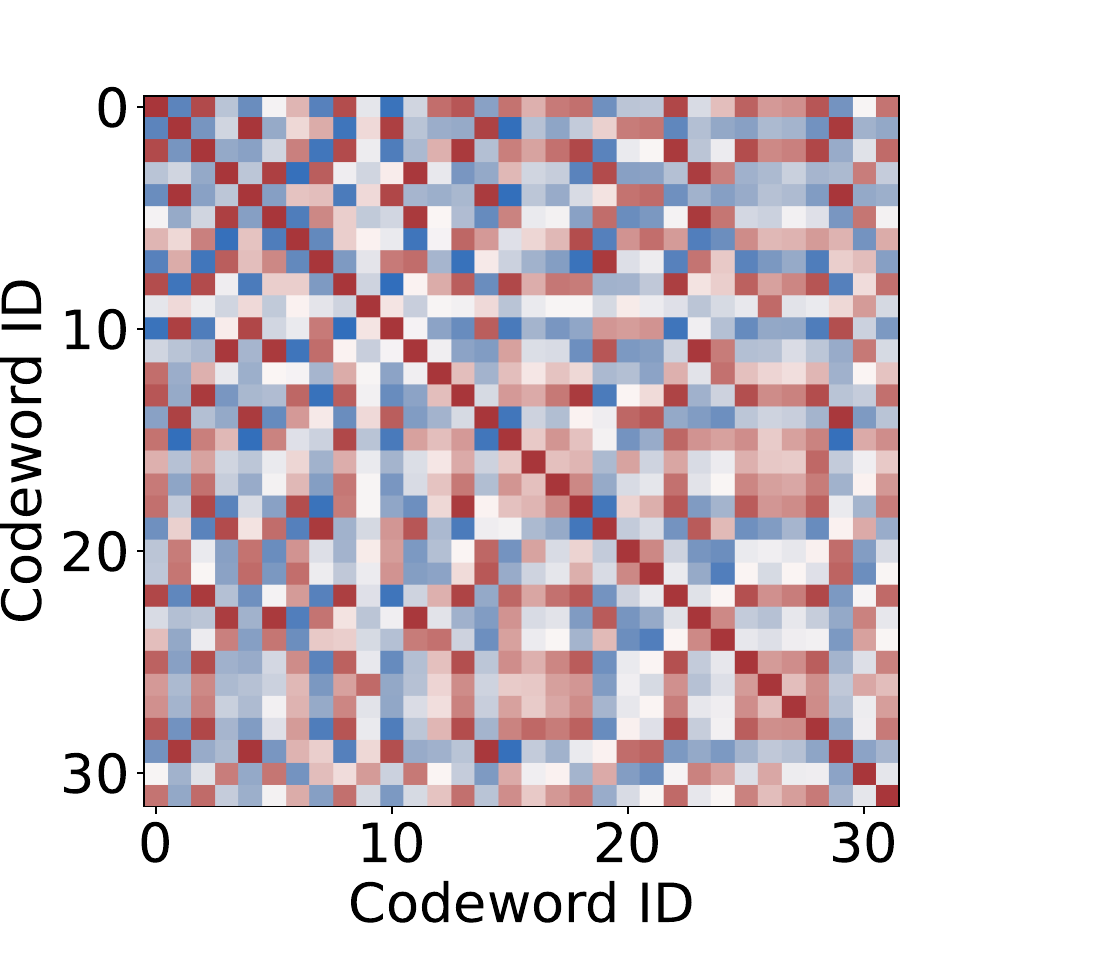}\label{fig:rq-vae}}
\caption{ Codeword similarity heatmaps for RQ-KMeans and RQ-VAE.
The off-diagonal correlations suggest that different SID codes preserve latent semantic proximity in the tokenizer-induced code space.}
\label{fig:appendix_codeword_heatmap}
\end{figure}

\label{app:codeword_structure}
Figure~\ref{fig:appendix_codeword_heatmap} provides additional codeword similarity heatmaps for RQ-KMeans and RQ-VAE. Similar to the OPQ result shown in the main text, both tokenizers exhibit non-trivial off-diagonal correlations, suggesting that learned SID code spaces preserve latent neighborhood relations among distinct codewords. These results indicate that the structural mismatch discussed in the main text is not specific to a particular tokenizer: semantic tokenizers can learn structured code spaces, whereas standard generators consume the resulting integer SIDs as independent categorical symbols.

\begin{figure}[h]
\centering
\includegraphics[width=0.95\columnwidth]{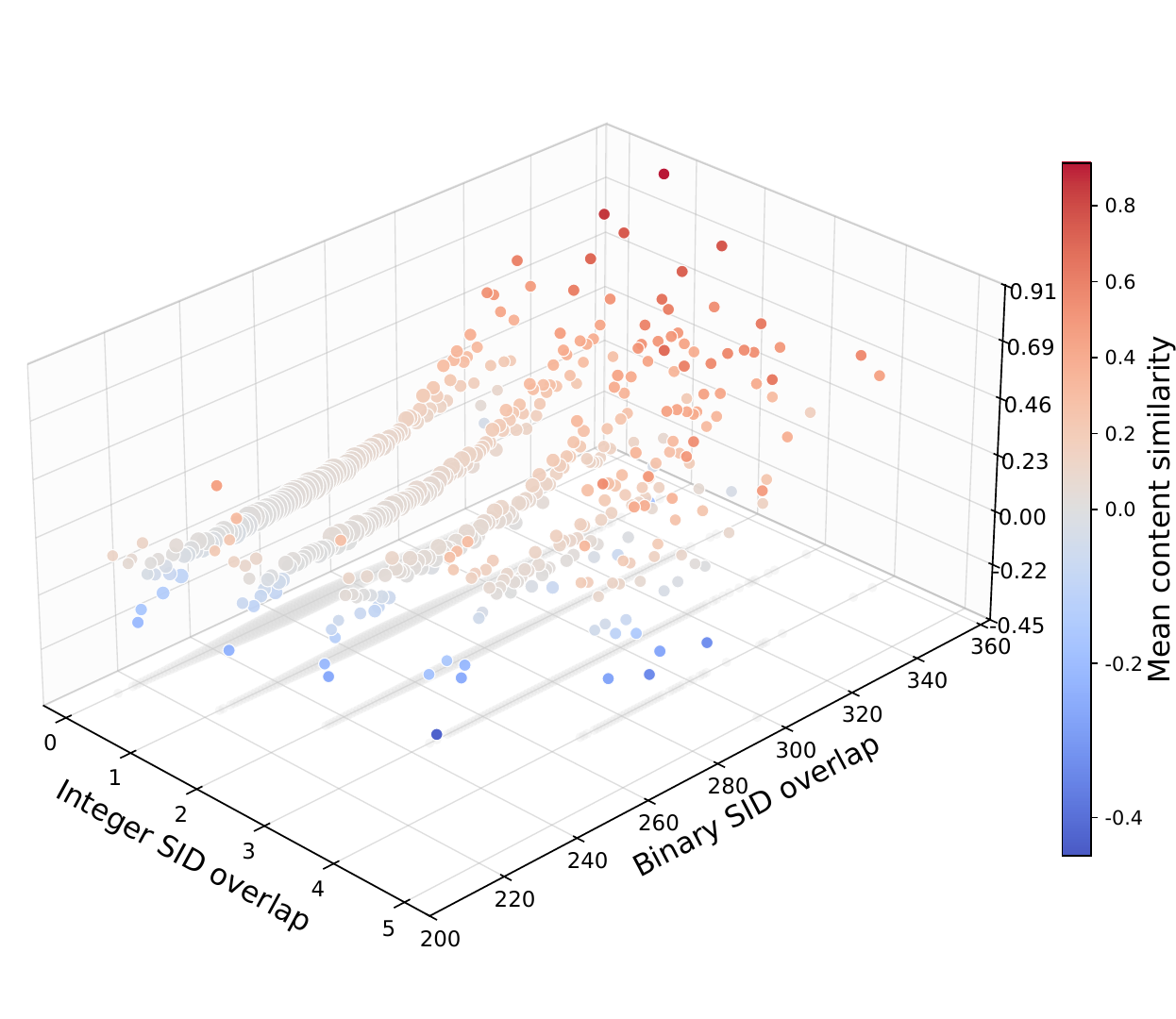}
\caption{Mean item content similarity with respect to integer SID overlap and Binary SID overlap. Only bins with sufficient item pairs are shown.}
\label{fig:bitsidoverlap}
\end{figure}

\begin{figure*}[t]
    \centering

    \begin{subfigure}

        \centering
        \includegraphics[width=0.85\linewidth]{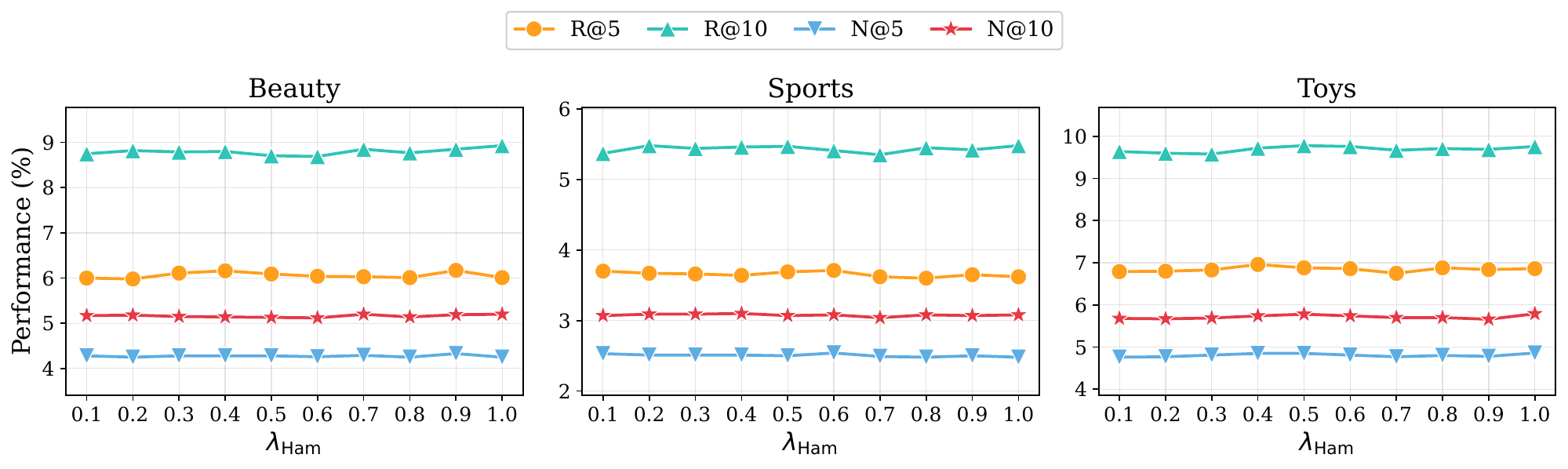}
            \vspace{-1.8em}
        \caption{Sensitivity analysis of $\lambda_{\mathrm{Ham}}$.}
        \label{fig:sensitivity_lambda_ham}
    \end{subfigure}
        \begin{subfigure}
        \centering
        \includegraphics[width=0.85\linewidth]{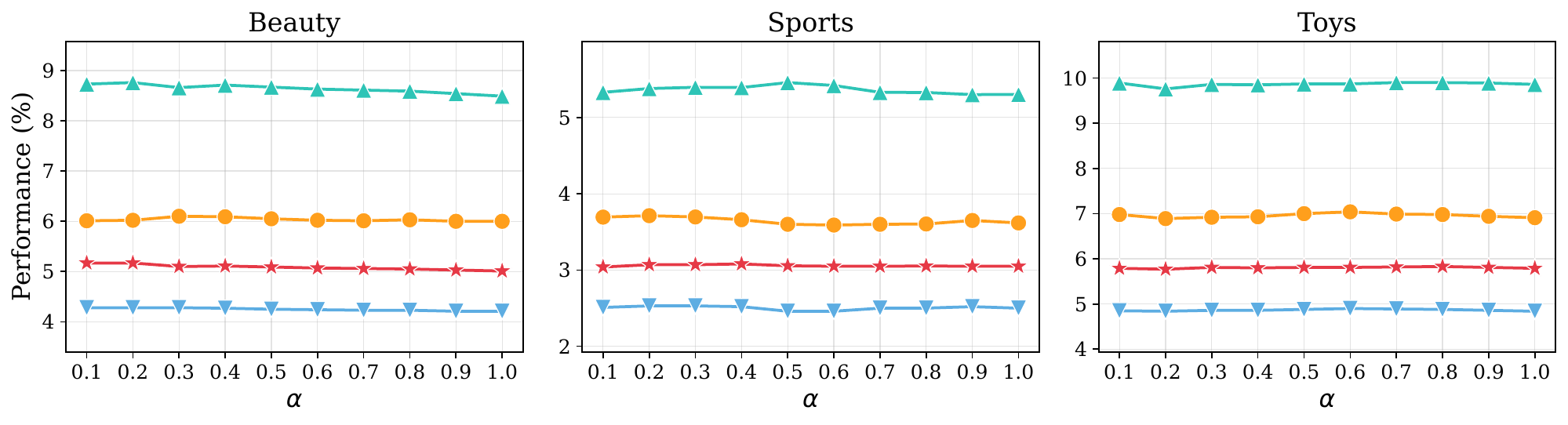}
            \vspace{-1.8em}

        \caption{Sensitivity analysis of $\alpha$.}
        \label{fig:sensitivity_alpha}
    \end{subfigure}
    \begin{subfigure}
        \centering
        \includegraphics[width=0.85\linewidth]{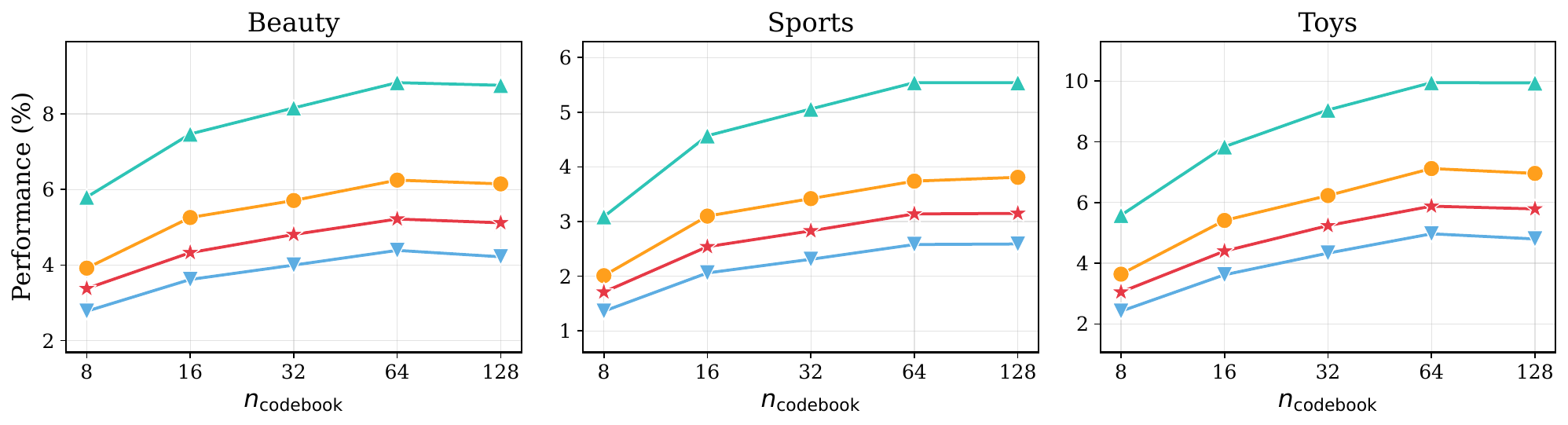}
            \vspace{-1.8em}

        \caption{Sensitivity analysis of $n_{\mathrm{codebook}}$.}
        \label{fig:sensitivity_n_codebook}
    \end{subfigure}

    \begin{subfigure}
        \centering
        \includegraphics[width=0.85\linewidth]{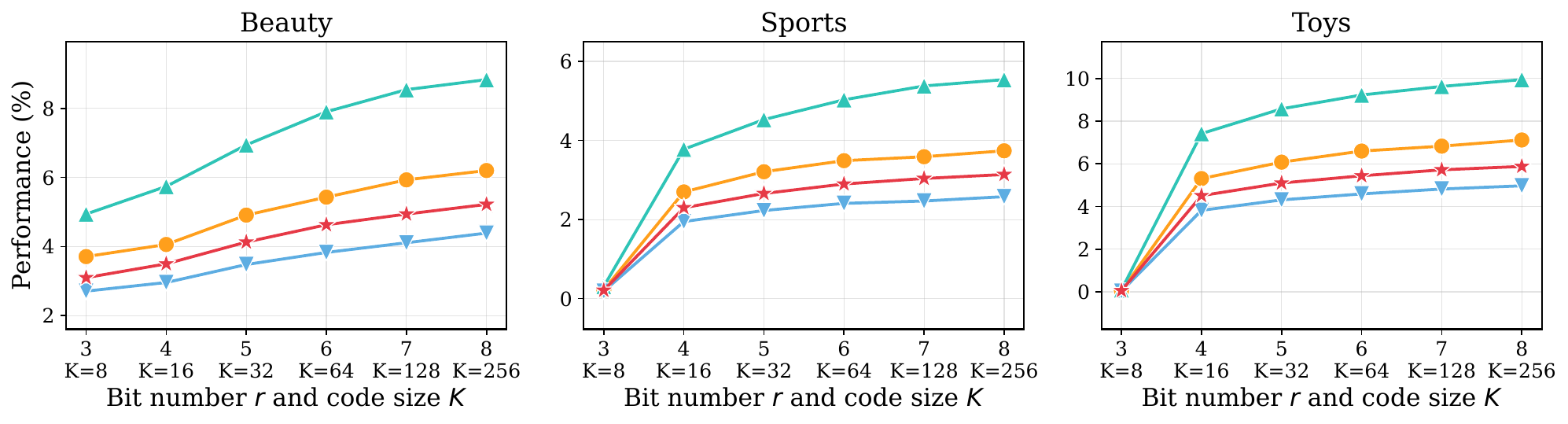}
            \vspace{-1.8em}
        \caption{Sensitivity analysis of bit number $r$.}
        \label{fig:bit_number}
    \end{subfigure}

    \label{fig:hyperparameter_sensitivity}
\end{figure*}

\begin{table*}[t]
\centering
\caption{Ablation results on Beauty, Sports, and Toys datasets.}
\label{tab:ablation}
\resizebox{\textwidth}{!}{
\begin{tabular}{lcccccccccccc}
\toprule
\multirow{2}{*}{Method} 
& \multicolumn{4}{c}{Beauty} 
& \multicolumn{4}{c}{Sports} 
& \multicolumn{4}{c}{Toys} \\
\cmidrule(lr){2-5} \cmidrule(lr){6-9} \cmidrule(lr){10-13}
& R@5 & R@10 & N@5 & N@10
& R@5 & R@10 & N@5 & N@10
& R@5 & R@10 & N@5 & N@10 \\
\midrule
\rowcolor{gray!15}
\multicolumn{13}{c}{\textbf{Binary SID Ablation}} \\
\midrule
OPQ\_SID 
& 0.0495 & 0.0701 & 0.0343 & 0.0409
& 0.0275 & 0.0423 & 0.0187 & 0.0234
& 0.0520 & 0.0719 & 0.0369 & 0.0433 \\
BDQ\_SID 
& 0.0497 & 0.0740 & 0.0344 & 0.0423
& 0.0297 & 0.0448 & 0.0202 & 0.0251
& 0.0524 & 0.0747 & 0.0367 & 0.0439 \\
Random SID + Binary 
& 0.0486 & 0.0671 & 0.0361 & 0.0421
& 0.0282 & 0.0396 & 0.0194 & 0.0231
& 0.0541 & 0.0706 & 0.0399 & 0.0453 \\
OPQ\_SID + Binary 
& 0.0566 & 0.0811 & 0.0404 & 0.0483
& 0.0364 & 0.0525 & 0.0248 & 0.0299
& 0.0641 & 0.0893 & 0.0445 & 0.0527 \\
\textbf{BDQ\_SID + Binary}
& \textbf{0.0609} & \textbf{0.0867} & \textbf{0.0432} & \textbf{0.0515}
& \textbf{0.0365} & \textbf{0.0538} & \textbf{0.0251} & \textbf{0.0307}
& \textbf{0.0674} & \textbf{0.0961} & \textbf{0.0467} & \textbf{0.0561} \\
\midrule
\rowcolor{gray!15}
\multicolumn{13}{c}{\textbf{Component Ablation}} \\
\midrule
w/o Basis Embedding (BE) 
& 0.0589 & 0.0859 & 0.0412 & 0.0499
& 0.0370 & 0.0546 & 0.0249 & 0.0306
& 0.0682 & 0.0979 & 0.0483 & 0.0579 \\
w/o Hamming-Consistent Reranking (HCR) 
& 0.0604 & 0.0871 & 0.0426 & 0.0512
& 0.0371 & 0.0546 & 0.0257 & 0.0313
& 0.0691 & 0.0977 & 0.0491 & 0.0583 \\
w/o Hamming Soft Target (HST) 
& 0.0614 & 0.0868 & 0.0430 & 0.0512
& 0.0370 & 0.0545 & 0.0255 & 0.0310
& 0.0688 & 0.0977 & 0.0487 & 0.0581 \\
\textbf{TopoGR} 
& \textbf{0.0620} & \textbf{0.0883} & \textbf{0.0439} & \textbf{0.0522}
& \textbf{0.0374} & \textbf{0.0554} & \textbf{0.0258} & \textbf{0.0314}
& \textbf{0.0712} & \textbf{0.0995} & \textbf{0.0497} & \textbf{0.0588} \\
\bottomrule
\end{tabular}
}
\end{table*}

\subsection{Content Similarity under Integer and Binary SID Overlap.}
\label{app:bit_content}
To further examine whether the binary topology reflects semantic relatedness in the original item content space, we analyze the mean content similarity of item pairs with respect to both integer SID overlap and Binary SID overlap. Here, integer SID overlap measures the number of shared SID tokens, while Binary SID overlap measures the number of matched bits in the corresponding Binary SIDs, which is equivalent to the complement of the full-SID Hamming distance. As shown in Figure~\ref{fig:bitsidoverlap}, item pairs with more shared integer SID tokens generally exhibit higher content similarity, confirming that exact SID overlap provides a coarse semantic signal. More importantly, within the same integer-overlap level, content similarity still varies substantially with Binary SID overlap. Item pairs with larger Binary SID overlap tend to have higher content similarity, even when their integer SID overlap is small or zero. This indicates that the Hamming geometry exposed by Binary SIDs captures fine-grained semantic relations that are not reflected by exact SID-token matching alone.

These observations are consistent with the controlled Hamming-distance analysis in Table~\ref{tab:hamming_close_far_n10} and Figure~\ref{fig:hamming_effect}. They further support our central claim that item relatedness resides not only in shared integer SID tokens, but also in the latent proximity among distinct SID codes. By making this topology explicit, TopoGR can exploit semantic relations among non-overlapping but structurally close SIDs during input modeling, training supervision, and inference-time reranking.

\subsection{Hyperparameter Sensitivity Analysis}
\label{app:hyperparam}

We conduct sensitivity analysis on three key hyperparameters of TopoGR across Beauty, Toys, and Sports datasets. Unless otherwise stated, all other hyperparameters are fixed at their default values.

\para{Effect of Hamming Loss Weight $\lambda_{\text{Ham}}$. }
Figure~\ref{fig:sensitivity_lambda_ham} shows the sensitivity of TopoGR to the Hamming soft-target loss weight $\lambda_{\text{Ham}}$ (Eq.~\ref{eq:hamming_kl}), which controls the relative contribution of topology-aware supervision versus standard cross-entropy. We observe that performance remains relatively stable across $\lambda_{\text{Ham}} \in [0.1, 0.5]$ on all three datasets, indicating that TopoGR is not overly sensitive to this hyperparameter. When $\lambda_{\text{Ham}}$ is too small (approaching 0), the model degenerates to standard categorical prediction without topology-aware smoothing, losing the benefit of Hamming soft targets. When $\lambda_{\text{Ham}}$ is too large (approaching 1.0), the Hamming proximity objective dominates training, potentially over-smoothing the prediction distribution and weakening the model's ability to discriminate the exact target code. The default value $\lambda_{\text{Ham}}{=}0.1$ achieves consistently strong performance across datasets, providing a suitable balance between exact token prediction and topology preservation.

\para{Effect of Reranking Weight $\alpha$. }
Figure~\ref{fig:sensitivity_alpha} presents the sensitivity to the reranking weight $\alpha$, which controls the strength of Hamming-consistent reranking (Eq.~\ref{eq:pred_sid_sim}) during inference. Specifically, $\alpha$ scales the Hamming similarity correction applied to the raw likelihood scores of candidate items. A larger $\alpha$ places more emphasis on Hamming proximity when re-scoring candidates, while $\alpha{=}0$ disables reranking entirely. The results show a clear inverted-U trend: performance peaks at moderate values ($\alpha \in [0.3, 0.5]$) and degrades at both extremes. At very small $\alpha$ (e.g., 0.1), the reranking correction is negligible, and the model relies almost entirely on raw likelihood scores without exploiting the Hamming structure of the Binary SID space. At very large $\alpha$ (e.g., 1.0), the Hamming proximity term dominates the final score, causing the model to favor candidates that are bit-level close to the prediction but may not be the true target—effectively introducing a proximity bias that overrides the learned generative signal. The optimal range $\alpha \in [0.3, 0.5]$ ensures that Hamming reranking provides a meaningful correction to disambiguate candidates with similar likelihood scores, without overwhelming the primary prediction.

\para{Effect of SID Length $n_{\text{codebook}}$. }Figure~\ref{fig:sensitivity_n_codebook} investigates the effect of SID length $M$ by varying it over $\{8,16,32,64,128\}$ while fixing the number of bits per position. Overall, recommendation performance consistently improves as $M$ increases across datasets and metrics. Longer SIDs provide greater representational capacity, enabling the tokenizer to encode more fine-grained item information and yielding richer binary features and Hamming relations for topology-aware modeling. Moreover, TopoGR predicts all SID positions in parallel, avoiding the serial decoding overhead commonly associated with long autoregressive SIDs. Nevertheless, increasing $M$ enlarges the input dimension and computational cost, and thus should be selected by balancing effectiveness and efficiency.

\para{Effect of the bit number $r$. } 
Figure~\ref{fig:bit_number} shows the effect of the bit number $r$, where the local code size is $K=2^r$.
The performance consistently improves as $r$ increases from 3 to 8.
When $r$ is too small, the local code capacity is insufficient and the induced Hamming space becomes overly coarse, leading to inferior recommendation performance.
A moderate bit number, e.g., $r=8$, provides sufficient code capacity and more fine-grained Hamming topology.




\subsection{Details of Hamming-close/Far Grouping}
\label{app:hamming_grouping}

We describe how the Hamming-close and Hamming-far groups are constructed.
For each test instance, let $v^\ast$ denote the target item and
$\mathcal{H}_u$ denote the user's historical items. Each item $v$ has an SID
$\mathbf{s}_v=[s_v^1,\ldots,s_v^M]$, where $M$ is the SID length. We first
compute the exact SID overlap count between the target item and each historical
item:
\begin{equation}
    \mathrm{Overlap}(v_h, v^\ast)
    =
    \sum_{m=1}^{M}
    \mathbb{I}[s_{v_h}^{m}=s_{v^\ast}^{m}].
    \nonumber
\end{equation}
For each test instance, we then take the maximum overlap count over the whole
history:
\begin{equation}
    O_u =
    \max_{v_h\in\mathcal{H}_u}
    \mathrm{Overlap}(v_h, v^\ast).
    \nonumber
\end{equation}
The test instances are grouped by this maximum overlap count, i.e.,
$O_u\in\{0,1,2,3,4,\geq5\}$. Thus, for buckets $0$--$4$, all instances in the
same bucket have exactly the same number of overlapped SID digits between the
target item and its closest-overlap historical item.

To define Hamming distance, we convert each SID digit into a fixed-length binary
code and concatenate all digits into a binary SID representation
$\mathbf{b}_v$. The Hamming distance between a historical item $v_h$ and the
target item $v^\ast$ is
\begin{equation}
    d_H(v_h,v^\ast)
    =
    \sum_{\ell}
    \mathbb{I}[b_{v_h}^{\ell}\neq b_{v^\ast}^{\ell}].
    \nonumber
\end{equation}
A smaller Hamming distance means that two items are closer in the binary SID
topology.

For the close/far comparison in Table~\ref{tab:hamming_close_far_n10}, we do
not use the minimum Hamming distance over all historical items directly.
Instead, we only consider historical items that achieve the maximum exact
overlap $O_u$. Specifically, we define
\begin{equation}
    D_u =
    \min_{v_h\in\mathcal{H}_u:
    \mathrm{Overlap}(v_h,v^\ast)=O_u}
    d_H(v_h,v^\ast).
    \nonumber
\end{equation}
This quantity measures the smallest Hamming distance among the historical items
that already have the maximum exact SID overlap with the target item.

Within each overlap-count bucket, we compute the empirical 30th and 70th
percentiles of $D_u$, denoted as $Q_{30}$ and $Q_{70}$. We then define
\begin{align}
    \text{Hamming-close}:&\quad D_u \leq Q_{30}, \nonumber\\
    \text{Hamming-far}:&\quad D_u \geq Q_{70}.
    \nonumber
\end{align}
The comparison focuses on the two clearly separated groups. Importantly, the
percentile thresholds are computed separately within each overlap-count bucket
and each dataset.

\section{Why Learned Hamming Geometry Works}
\label{appendix:hamming_geometry}

This section provides a simple justification for why Hamming distance is meaningful in the learned Binary
SID space.
Binary SID is not an arbitrary binary encoding; instead, it is learned through a bit-compositional reconstruction structure.

For the $m$-th SID position, the bit-compositional representation of item $v$ is defined as
\[
\mathbf{g}_v^m=\sum_{l=1}^{r} b_v^{m,l} \mathbf{a}_{m,l},
\]
where $b_v^{m,l}\in\{-1,+1\}$ denotes the $l$-th binary bit and $\mathbf{a}_{m,l}$ is the corresponding learnable bit basis.
For two items $u$ and $v$, we have
\[
\mathbf{g}_u^m-\mathbf{g}_v^m
=
\sum_{l=1}^{r}
(b_u^{m,l}-b_v^{m,l})\mathbf{a}_{m,l}.
\]
Since identical bits have zero difference, only the differing bits contribute to the representation difference:
\[
\mathbf{g}_u^m-\mathbf{g}_v^m
=
\sum_{l:b_u^{m,l}\neq b_v^{m,l}}
(b_u^{m,l}-b_v^{m,l})\mathbf{a}_{m,l}.
\]
Taking the $\ell_2$ norm and using $|b_u^{m,l}-b_v^{m,l}|=2$ for different bits, we obtain
\[
\|\mathbf{g}_u^m-\mathbf{g}_v^m\|_2
\le
2
\sum_{l:b_u^{m,l}\neq b_v^{m,l}}
\|\mathbf{a}_{m,l}\|_2.
\]
This shows that two codes with small Hamming distance share most learned bit bases and differ only in a few basis components.

For the full SID, define the set of differing bit positions as
\[
\Delta(u,v)=\{(m,l)\mid b_u^{m,l}\neq b_v^{m,l}\}.
\]
The full-SID Hamming distance is then
\[
d_H(B_u,B_v)=|\Delta(u,v)|.
\]
Therefore, the Hamming distance can be interpreted as counting how many learned basis components are changed between two Binary SID codes.

This provides a structural explanation for why Hamming distance works in our method.
Since the tokenizer is trained to reconstruct item representations from these bit-basis compositions, the learned Hamming geometry is tied to the reconstruction structure rather than being an arbitrary distance over binary strings.
This also explains why random binary SIDs are less effective: although they also have Hamming distances, their bits are not associated with reconstruction-trained basis components and thus are not aligned with item representation reconstruction.

\end{document}